\documentclass[aps,prl,twocolumn,superscriptaddress,nofootinbib,notitlepage,longbibliography]{revtex4-2}
\usepackage{bm}
\usepackage{amsmath}
\usepackage{amssymb}
\usepackage{mathdots}
\usepackage{graphicx}
\usepackage{hyperref}
\usepackage{xcolor}
\usepackage{multirow}
\usepackage{dsfont}

\begin{document}

\title{Superconductivity from Quasiparticle Pairing of Intervalley Coherent State in Rhombohedral Trilayer Graphene}

\author{Chun Wang Chau}
\affiliation{Department of Physics, Hong Kong University of Science and Technology,
Clear Water Bay, Hong Kong, China}
\affiliation{Theory of Condensed Matter Group, Cavendish Laboratory, University of Cambridge, J.\,J.\,Thomson Avenue, Cambridge CB3 0HE, UK}

\author{Shuai A. Chen}
\email{chsh@ust.hk}
\affiliation{Department of Physics, Hong Kong University of Science and Technology,
Clear Water Bay, Hong Kong, China}

\author{K. T. Law}
\email{phlaw@ust.hk}
\affiliation{Department of Physics, Hong Kong University of Science and Technology,
Clear Water Bay, Hong Kong, China}

\date{\today}

\begin{abstract}
    Superconductivity is observed in rhombohedral trilayer graphene in a narrow regime between the flavor-symmetric state and the symmetry breaking phase,
    which cannot be described by the conventional Bardeen–Cooper–Schrieffer theory. The measured coherence length, for instance, is roughly two orders of magnitude shorter than the value predicted by the Bardeen–Cooper–Schrieffer relation based on the large fermi velocity and an extremely low charge carrier density of the flavor-symmetric phase. 
    To resolve the discrepancies, we propose that the rhombohedral trilayer graphene superconducting phase arises from the pairing of quasiparticles of the adjacent inter-valley coherent state.
    We illustrate the superconducting phenomenology using gapped Dirac cones with the chemical potential $\mu$ close to the valence band's edge. 
    Our findings indicate that the transition temperature $T_c$ obeys $T_c\propto \epsilon_D\exp(-2/\rho_\mathrm{qp}U)$ with the density of states $\rho_\mathrm{qp}$ of intervalley coherent state quasiparticles, which is much suppressed compared to predictions from the Bardeen–Cooper–Schrieffer theory.
    The coherence length $\xi$ we predict behaves according to $\xi\sim v/\sqrt{\mu T_c }$ with $v$ being the velocity of Dirac cone.
    Applying our assumption to a microscopic model, our predictions align well with experimental data and effectively capture key measurable quantities such as the transition temperature $T_c$ and the coherence length $\xi$ without parameter fine-tuning. 
\end{abstract}  

\maketitle

\begin{figure*}[ht!]
	\includegraphics[width=2\columnwidth]{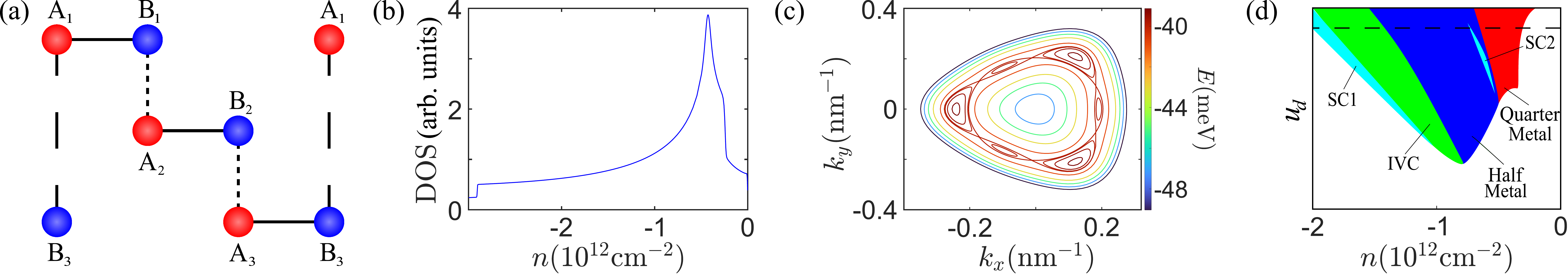}
    \caption{
    Lattice structure, dispersion spectrum, and phase diagram of RTG.
    (a) Schematic illustration of the lattice structure of RTG.
    Note that only $A_1$ and $B_3$ do not have direct interlayer hopping, and hence are of lower energy in comparison with other sites. 
    As such at the low charge carrier density, they act as active sites of the system.
    (b)  The density of states (DOS) at the valence band of electrons in a flavour-symmetric phase.
    There is a jump at charge carrier density $n\approx-2.9\times10^{12}\mathrm{cm}^{-2}$, indicating a transition to an annulus Fermi surface from a complete one. 
    Also, there is a Van Hove singularity when $n\approx-0.5\times10^{12}\mathrm{cm}^{-2}$, corresponding to Lifshitz transition to disconnected pockets. 
    (c) Contour plot of the band structure at K valley. 
    When the chemical potential varies, the Fermi surface topology is changed from disconnected pockets to annulus, and then to a completed surface, giving rise to the density of states in (b).
    (d) The phase diagram from the experiment in Ref.~\onlinecite{SC-RTG}.
    The flavour-symmetric state is left uncoloured.
    The SC1 occurs in between the flavour-symmetric state and the IVC state in a very narrow window of charge carrier densities.  
    In obtaining (b) and (c),  we set the potential difference between the outer layers of RTG as $u_d=34.5$meV, and the horizontal dashed line in (d) corresponds to $u_d=34.5$meV.
    } \label{fig:1}
\end{figure*}

\section{Introduction}

    Graphene heterostructures, in particular twisted bilayer graphene, have been observed experimentally to host various strongly correlated phases \cite{CI_MAG,AH_TBG,EF_TBG,MAG+WSe,SC_MAG,SC+CS_MAG,SC_TBG,TS_TBG,NP_TBG,CO_TBG,MBC_TBG}. However, many of these heterostructures are difficult to realize in experiments due to structural instability \cite{SI_TBG,MI_TBG}. Recently, breakthrough \cite{SC-RTG, HM-RTG, IVC-Exp} has been made in a more stable graphene-based system, namely ABC-stacked rhombohedral trilayer graphene \cite{BS, TW-RTG} (RTG), and has opened up exciting opportunities for studying strongly correlated phases of matter \cite{SOM-RTG, ECP-RTG, CP-RTG, EC-RTG}. By adjusting the electron density and applying electric fields perpendicular to the material \cite{BG-TG}, researchers can customize and investigate these states in RTG.

    One particularly intriguing development is the discovery of superconductivity in RTG, which manifests in two distinct phases referred to as SC1 and SC2 in \cite{SC-RTG}. SC1 occurs within the flavour-symmetric phase and exhibits a maximum critical temperature of $T_{c1} \sim 100 \mathrm{mK}$ and superconducting coherence length $\xi\sim200\mathrm{nm}$. There are significant discrepancies when applying the BCS theory. 
    First, to align with experimental observations, the transition temperature $T_c$ should obey $T_c\propto\epsilon_D\exp(-2/\rho U)$. It deviates from the antiadiabatic limit of Bardeen-Cooper-Schrieffer (BCS) theory applicable to low density of states \cite{SC-RTG,2016PNAS..113.4646G} $T_c\propto\epsilon_D\exp(-1/\rho U) $ where $\rho$ is the density of states at fermi energy and $U $ is the strength of attractive interaction.  
    Second, the coherence length reported in experiments is two orders of magnitude shorter than the value predicted by the BCS relation $\xi= \hbar v/T_c$, provided the large Fermi velocity $v$ of Dirac cone within graphene heterostructures. 
    Several theories have been proposed to elucidate the electrons' pairing mechanism behind SC1 in RTG, such as electron-phonon coupling \cite{AP-RTG, P-RMG}, Kohn-Luttinger-like mechanism \cite{RG-IVC, AFS-RTG, KT-RTG, SO-RTG}, direct coupling mediated by Coulomb interaction \cite{RGA-RTG, FRG-RTG, MC-RTG}, and pairing facilitated by the proximity to a correlated metal \cite{DB3-RTG, IVC, SR-RTG}. 
    However, the influence of the neighbouring symmetry-breaking phase, which is likely an intervalley coherent (IVC) state, on the properties of the superconducting states remains unclear, in particular  when it comes to explaining experimental measurements.

    In this paper, we propose that SC1 observed in RTG is due to direct pairing between quasiparticles of IVC state instead of the pairing between the bare electrons. 
    We first clarify clear discrepancies from the experimental results \cite{SC-RTG} both in the transition temperature and upper critical field, when we assume direct pairing between electrons. 
    Instead, when in close vicinity with the phase transition from IVC to a flavour symmetric state, where the chemical potential cuts the band edge of the lower energy IVC quasiparticle band, there is an increase in density of state for the IVC quasiparticles. This favours direct pairing between IVC quasiparticles, leading to the IVC-SC phase, where the IVC order parameter remains intact. We then analytically study the unique properties of superconductivity near the band gap using a toy model that captures the key feature of the IVC state, namely, nearly massless Dirac fermions located at $K$ and $K'$ valleys. By parameterizing the toy model accordingly to the RTG, we approximate the coherence length that is similar in value to what is observed in experiments, and explain a narrower window of superconductivity as opposed to the BCS theory \cite{BCS} on a highly dispersive band.

    In closing the paper, we perform numerical calculations explicitly on the IVC states of RTG, calculating the transition temperature and upper critical field, which match well and capture key features of the experimental result. 
    We also commented on the possible implication of the quantum metric near the phase boundary of the SC and the IVC state. 

\begin{figure*}[t!]
    \includegraphics[width=2\columnwidth]{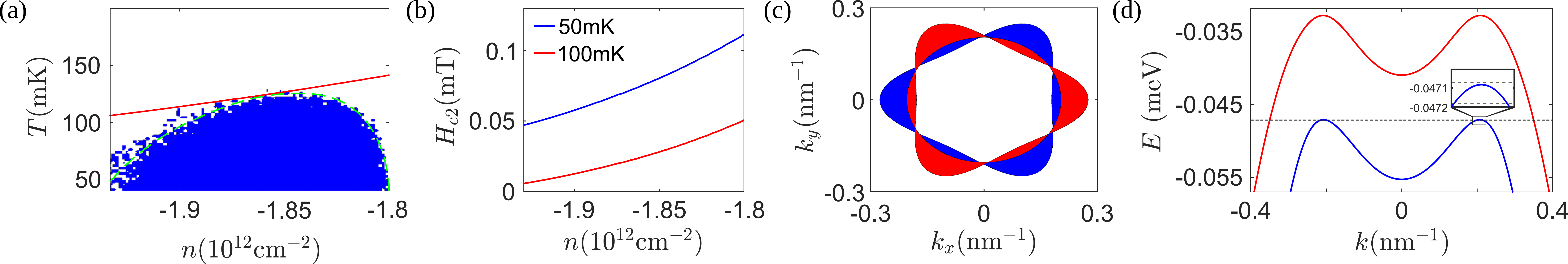}
    \caption{
    Predictions from the theory by assuming SC1 from flavour-symmetric phase.
    (a) The mean-field transition temperature (red curve) by assuming superconductivity from a flavour-symmetric phase. The blue region is extracted for SC1  from the experiment in Ref.~\onlinecite{SC-RTG}, and the green dashed line indicates the corresponding transition temperature. The predictions deviate from the experimental measurements in particular at the phase boundary with charge carrier density $n=-1.8\times10^{12}\mathrm{cm}^{-2}$. 
    (b) Out-of-plane upper critical field $H_{c2}$ by assuming superconductivity from a flavour-symmetric phase.
    The two curves correspond to temperature $T=50$mK (Blue) and $T=100$mK (Red). 
    The predictions are of two orders of magnitude smaller than the experimental results $\sim10$mT in Ref.~\onlinecite{SC-RTG}.
    Within the window of SC, the upper critical field is predicted to increase monotonically, as contrast to the experimental observations where $H_{c2}$ reach a maximum at $n\approx-1.82\times10^{12}\mathrm{cm}^{-2}$ and decreases to $0$ at $n\approx-1.8\times10^{12}\mathrm{cm}^{-2}$.
    (c) Fermi surfaces corresponding to the Lifshitz transition from the annulus to isolated pockets, when $u_{d}=34.5$meV at $n=-0.5\times10^{12} \mathrm{cm}^{-2}$.
    The red and blue regions correspond to the valence band at K and K$^\prime$ valleys, respectively.
    The Fermi surfaces exhibit nearly perfect nesting, thus having strong instability under repulsive interaction.
    (d) Dispersion spectrum of the quasiparticles in Eq.~\eqref{eq:qpdis} in the IVC state at different momenta $k=\sqrt{k_x^2+k_y^2}$. 
    The dashed lines enclose the chemical potential of a very narrow regime where a superconducting state SC1 is observed in experiments, and the inset zooms in to show the regime.
    }
    \label{fig:2}
\end{figure*}

\section{Results}

\subsection{Discrepancies in Superconductivity of RTG}
\label{sec:SC_N}

For RTG, where the lattice structure is illustrated in Fig.~\ref{fig:1}(a),
superconducting state SC1 is observed within the window of charge carrier density $n\approx-1.93\times10^{12}\mathrm{cm}^{-2}$
to $-1.8\times10^{12}\mathrm{cm}^{-2}$ when $u_{d}$, which is the potential difference between the outer layers of RTG \cite{BS}, is approximately $34.5$meV in experiment \cite{SC-RTG}. Despite the increase in density of states as illustrated in Fig.~\ref{fig:1}(b)
due to the trigonal warping \cite{TW-RTG} of the band structure as shown in Fig.~\ref{fig:1}(c), when we further decrease the carrier
density, competing states, for example, the spin polarized state \cite{HM-RTG}, are more energy favourable, and thus a superconducting
state is not observed, until SC2 \cite{SC-RTG}, which is deep within
the half-metal regime. A schematic for the phase diagram is illustrated in Fig.~\ref{fig:1}(d). Focusing on SC1, at $n\approx-1.85\times10^{12}\mathrm{cm}^{-2}$,
corresponding to chemical potential of $\mu\approx-43.4\mathrm{meV}$, the transition temperature takes the maximal value of $T_c\approx120\mathrm{mK}$.
By fitting the maximum transition temperature at the specific charge carrier density, we estimate an attractive interaction strength of approximately $U \approx 4 \mathrm{eV}$ using the self-consistency equation in a mean-field theory, as discussed in detail in Sec. Method. Specifically, the dispersion spectrum is extracted from the 6-band Hamiltonian for RTG \cite{BS, HM-RTG}, detailed in Supplementary Material.
Indeed, when considering electron pairing in the flavour-symmetric state and examining a wide range of charge carrier densities, a noticeable discrepancy emerges between the predicted trend of the transition temperature according to the mean-field theory and the experimental observations illustrated in Fig.~\ref{fig:2}(a). This discrepancy highlights the inadequacy of the consideration of pairing in the flavour-symmetric state in capturing the intricacies of the superconducting phase.

Significant contradiction from the experimental result can also be observed when we
attempt to calculate the superconducting coherence length.  At charge carrier density $n\approx-1.85\times10^{12}\mathrm{cm}^{-2}$, where the transition
temperature is maximized, under BCS relation \cite{BCS}, the coherence length can be approximated by $\xi_0\sim0.18\hbar v/k_BT_c$, where $v$ is the typical velocity of graphene heterostructure. We note the chemical potential corresponding to the experimental regime is energetically far away from the $K(K')$ points, thus the interlayer tunnelling effect is negligible. Therefore, the typical velocity can be approximated by the fermi velocity of massless graphene monolayer $\hbar v\approx6.6\times10^{-1}$eVnm.
Using $T_c\approx120$mK, we can approximate the coherence length as $\sim10^4$nm, which is roughly
two orders of magnitude larger than the coherence length ($150-250$nm)
extracted from the experiment using the perpendicular upper critical
field $H_{c2}={\phi_{0}}/{2\pi\xi^{2}}$
where $\phi_{0}={h}/{2e}$ is the superconducting quantum flux. If we additionally account for the low charge carrier density, from numerical calculations, at the base temperature
($\approx50$mK) of the experiment, the coherence length lies between $1650-1800$nm, which is still an order of magnitude larger than the experimental result. 
A plot of $H_{c2}$ from numerical calculation is shown in Fig.~\ref{fig:2}(b).

These discrepancies from experiments to BCS relation are clear indications that the superconductivity SC1 observed in RTG may not originate from the flavour-symmetric state, namely pairing of bare
electrons. Instead, SC1 might emerge from the neighbouring competing state,
which is referred to as the partially isospin polarized (PIP) state
\cite{SC-RTG,HM-RTG} and most likely corresponds to the intervalley
coherence state (IVC) \cite{RG-IVC,IVC}.

In the following, we propose that Cooper pair in SC1, instead of originating from pairing between electrons, is from the direct pairing of quasiparticles of the IVC state, thus referred to as IVC-SC in the remainder of this paper. Therefore, SC1 is characterized by both the IVC order parameter and the superconducting order parameter simultaneously. 
In addition, we point out that the occurrence of IVC-SC in proximity to the phase boundary between the IVC state and the flavour-symmetric state is not merely coincidental but rather a key characteristic. 

\subsection{Quasiparticle pairing in gapped Dirac cones}
In this section, we will study the superconducting phase using a toy model. 
In the experimental regime, consideration of IVC quasiparticle (see Sec.~Method) pairing allows realization of superconductivity in the vicinity of the band gap, and the properties of the superconducting phase are distinct from the standard BCS theory.
In our toy model, to mimic the IVC state in RTG, we use two Dirac
cones, of Fermi velocity $v$, centred at the $K$ and $K^\prime$
valley, respectively. By acting with an external potential, for example
a displacement field, we open up a mass gap $m$ between the originally
touching Dirac bands. 
The IVC phase can be effectively described in the
continuum limit by a massive Dirac fermion Hamiltonian $H=H_0+V_\mathrm{IVC}$:
\begin{align}
H_0 & =\sum_{s\tau} \psi_{\tau s}^\dagger(\mathbf q) h_\tau(\mathbf q) \psi_{\tau s}(\mathbf q),
\label{eq:hs1} \\
V_\mathrm{IVC} &= \sum_{s} \psi_{+ s}^\dagger(\mathbf q) \Delta_{\mathrm{IVC}}(\mathbf q) \psi_{- s}(\mathbf{q})+h.c.
\label{eq:hs2}
\end{align}
with the spinor $\psi_{\tau s}= [a_{\tau sA},a_{\tau sB}]^T$ at two sublattices $\lambda = A,B$ given the spin index $s=\uparrow,\downarrow$ and the valley index $\tau = \pm$.
In specific, $h_\tau(\mathbf{q})$ describes  massive Dirac cone and is given by:
\begin{align}
h_{+}(\mathbf{q}) & = v\mathbf{q}\cdot(\tau\sigma_{x}\hat{\mathbf{e}}_{x}+\sigma_{y}\hat{\mathbf{e}}_{y})+m\sigma_{z},
\end{align}
where $\sigma_{x,y,z}$ correspond to the Pauli matrices defined in the sublattice basis. $V_\mathrm{IVC}$ introduces the IVC order parameter which for simplicity is assumed uniform and real (see Methods for details). {Below we will use the minimalistic toy model of massive Dirac cone to obtain an analytical closed form, in order to qualitatively understand the phenomenology of band edge superconductivity.}

There are four quasiparticle bands in Eq.~(\ref{eq:hs1}-\ref{eq:hs2}) given the spin index. In particular, we will focus on the quasiparticle band of the dispersion spectrum \cite{IVC} $E(\mathbf q)=-\sqrt{v^{2}q^{2}+m^{2}}-\Delta_{\mathrm{IVC}}$, and
consider only s-wave on-site effective attractive interaction:
\begin{equation}
H_{\mathrm{int}}=-U\sum_{s\lambda\lambda^\prime}\int d^{2}\mathbf{r}\ a_{+s\lambda}^{\dagger}(\mathbf{r})a_{-\bar s\lambda^\prime}^{\dagger}(\mathbf{r})a_{-\bar s\lambda^\prime }(\mathbf{r})a_{+s\lambda}(\mathbf{r}),
\end{equation}
where $U$ is attractive interaction strength, and  $\bar s$ indicates opposite spin. 
We do comment the possibility of unconventional pairing \cite{AFS-RTG, SO-RTG, RGA-RTG, FRG-RTG}, given the interaction is non-local. 
We can analyse critical temperature and coherence length of the resulting SC though the both mean field theory and the Ginzburg-Landau theory (see Sec.~Method).

We focus on the case where chemical potential $|\mu|-m<\epsilon_D$, which is the relevant regime to experiments, with $\epsilon_D$ represents the Debye energy for the attractive interaction.
For the regime where the chemical potential satisfies $\beta_{\mathrm{MF}}(m-|\mu|)\gg1$,
thus away from the mass gap, the conventional contribution dominates, by a factor of $\sim\beta_{\mathrm{MF}}|\mu|\gg1$.
Thus, 
the system can be treated as conventional when away from the band edge, where 
we can retrieve results similar to the BCS limit, as detailed in Sec.~Method

We now consider the case where the chemical potential satisfies $\beta_{\mathrm{MF}}|m+\mu|\ll1$, thus in proximity with band gap.
The mean-field temperature is solved from the linearized gap equation in  Eq.~\eqref{eq:TMF}:
\begin{equation}
T_{\mathrm{MF}}\propto\epsilon_{D}\exp\left(-\frac{2}{\rho_\mathrm{qp} U'}\right),
\label{eq:T_b}
\end{equation}
where $\rho_\mathrm{qp}$ is the density of states of the quasiparticles at the fermi energy. We have also defined $\frac{1}{U'}=\frac{1}{U}-\frac{\epsilon_{D}\mathcal{A}}{4\pi v^{2}}$, with the correction term ${\epsilon_{D}\mathcal{A}}/{4\pi v^{2}}\ll1$.
Note that there is an extra factor of two in the exponential term in comparison
with the BCS limit, as no states exist within the band gap, effectively halving the integral range over energy when calculating mean-field temperature self-consistently. This hints at superconductivity being more suppressed
near the band edge, thus the range of doping where superconductivity
can be observed, is expected to be narrower when compared with the prediction
from BCS. The conventional coherence length in this regime is given
by:
\begin{equation}
\xi_{\mathrm{con}}\approx\frac{1}{4}\frac{v}{\sqrt{|\mu|T_{\mathrm{MF}}}}\left(\frac{T_{\mathrm{MF}}-T}{T_{\mathrm{MF}}}\right)^{-1/2},
\label{eq:l_c_b}
\end{equation}
which gives rise to the Ginzburg-Landau coherence length $\xi_\mathrm{GL,con}\approx\frac{1}{4}\frac{v}{\sqrt{|\mu| T_{\mathrm{MF}}}}$. 
This relation deviates from a BCS relation $\xi_\mathrm{BCS}\approx v_F/T_\mathrm{MF}$.
As for the contribution from quantum
metric,  when the chemical potential is at the band edge (i.e. $|\mu|=m$), we have:
\begin{equation}
   \xi_{\mathrm{qm}}\approx \frac{v}{4m}\left(\frac{T_{\mathrm{MF}}-T}{T_{\mathrm{MF}}}\right)^{-1/2}
    \sqrt{\gamma^2-1+2\ln\left(\frac{\beta_{\mathrm{MF}}^2m^2}{\pi}\right)},
\end{equation}
where $\gamma$ is the Euler's constant, and we have the quantum metric Ginzburg-Landau coherence length $\xi_\mathrm{GL,qm}\approx \frac{v}{2m}\ln(\beta_\mathrm{MF}m)$.
As such we can conclude that the quantum metric contribution is smaller
than, but can be comparable with the conventional coherence length, by a
factor of $\sim2\sqrt{\ln(\beta_{\mathrm{MF}}m)/\beta_{\mathrm{MF}}m}<1$
at the valence band edge.

Another intriguing scenario arises when the chemical potential is in proximity to the edge of the band, but lies within the band gap. At zero temperature, one would not typically anticipate a superconducting phase since the ground state is an insulator.
However, a SC phase can occur at a finite temperature due to a metallic phase induced by temperature fluctuations. Of course, the corresponding mean-field transition temperature gets suppressed. 
As the transitional temperature is small such that $\beta_\mathrm{MF}|\mu+m|\gg 1$, the conventional coherence length is proportional to:
\begin{equation}
    \xi_\mathrm{GL,con}\propto \frac{v}{T_\mathrm{MF}} e^{-\beta_\mathrm{MF}|\mu+m|/2},
\end{equation}
which goes to zero when we approach the phase boundary given a fixed chemical potential $\mu$. On the contrary, the quantum metric coherence length diverges and is proportional to:
\begin{equation}
    \xi_\mathrm{GL,qm}\propto \frac{v}{T_\mathrm{MF}}\sqrt{\frac{\ln(\beta_\mathrm{MF}m)}{\beta_\mathrm{MF}m}}.
\end{equation}
Obviously, $\xi_\mathrm{GL,qm}$ diverges when $T_\mathrm{MF}$ approaches zero. 
As such, without considering the effect of the quantum metric, we expect the Ginzburg-Landau coherence length to remain small as we approach the phase boundary. On the contrary, if the effect of quantum metric is included, we would instead predict a divergent Ginzburg-Landau coherence length.
As we will see, this case has potential implications for the interesting results in experimental measurements in RTG \cite{SC-RTG}.

\begin{figure*}[ht!]
    \includegraphics[width=2\columnwidth]{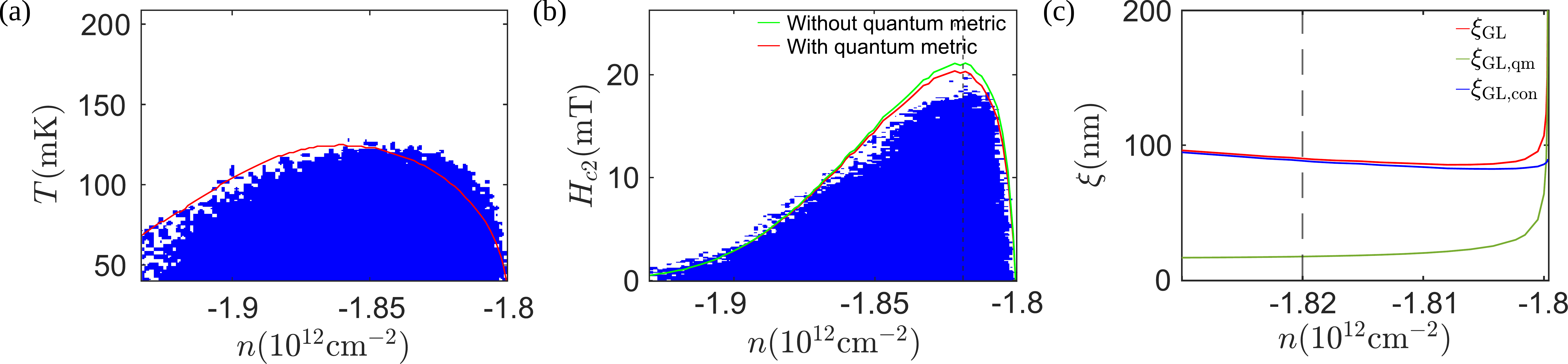}
    \caption{
    Predictions from the theory by assuming SC1 originates from pairing of quasiparticles in IVC phase. (a) the mean-field transition temperature, (b) upper critical field $H_{c2}$ at $T=50$mK, and (c) the Ginzburg-Landau coherence lengths $\xi$ at $n\approx-1.82\times10^{12}\mathrm{cm}^{-2}$.
    In (a), the SC1 occurs in the blue region which is extracted from the experiment in Ref.~\onlinecite{SC-RTG} with the transition temperature approximated by the boundary of the blue region.
    The mean-field temperature (red line) captures the key features of the experimental data, especially the disappearance of SC1 at charge carrier density $n\approx-1.8\times10^{12}\mathrm{cm}^{-2}$.
    In (b), the green curve corresponds to $H_{c2}$ only accounting for the conventional contribution, while the red curve additionally accounts for quantum metric.
    The black dashed line indicates the chemical potential exactly at the Dirac point of the quasiparticle band in IVC.
    The blue region is extracted from \cite{SC-RTG}.
    Our theoretical predictions capture both the maximal in $H_{c2}$ in close proximity with the band edge, and the sharp disappearance of superconductivity at $n\approx-1.8\times10^{12}\mathrm{cm}^{-2}$.
    In (c), the green and blue curves correspond to the quantum metric and conventional contributions respectively, and the red curve is the overall Ginzburg-Landau coherence length $\xi_\mathrm{GL}$.
    The conventional contribution $\xi_\mathrm{GL,con}$ plays a dominating role, and the quantum metric contribution $\xi_\mathrm{GL,qm}$ has a correctional effect in most regimes.
    However, near the phase transition from IVC-SC to IVC state of which the chemical potential lies between the band gap,
    the quantum metric contribution $\xi_\mathrm{GL,qm}$ increases significantly, surpassing the conventional contribution $\xi_\mathrm{GL,con}$ due to the vanishingly low density of states of the IVC quasiparticles at the fermi energy. 
    It leads to a divergence of coherence length $\xi_\mathrm{GL}$ at the phase boundary, in contrast to a finite value for the conventional contribution at $n\approx-1.8\times10^{12}\mathrm{cm}^{-2}$. 
    Here the calculations are based on the microscopic model for RTG in SM \cite{SM-TG}.
  }
    \label{fig:3}
\end{figure*}

\subsection{Prediction from toy model on SC in RTG} \label{sec:IVC-SC} 
{ With the toy model, we have obtained an analytical closed form that can invoke qualitative insight on the phenomenology of superconductivity observed in RTG. }
The IVC-SC phase is observed in the vicinity of the
phase boundary in between the IVC and the flavour-symmetric state, at which the
chemical potential cuts the band edge of the lower energy IVC band (coloured blue in Fig.~\ref{fig:2}d).
This results in a discrete jump in DOS for the IVC quasiparticle, and leads to strong instability towards a flavour symmetric state.
Since only quasiparticles from the edge of the lower energy IVC band majorly contribute to
the formation of Cooper pairs, the lack of states within the {direct band gap} at zero temperature is well captured 
by the proposed toy model. As such, physical properties of IVC-SC, for example, coherence
length, can be approximated using the result from the band edge superconductivity of massive
Dirac fermions. Given the effective mass gap of $m\approx34.5$meV
, with mean-field temperature $T_{\mathrm{MF}}=120$mK and graphene typical
velocity $\hbar v=6.6\times10^{-1}$eVnm, the coherence length
at base temperature $\approx50$mK can be approximated using Eq.~\eqref{eq:l_c_b}
to be $126$nm, which is very close to the value ($150$-$250$nm) reported in the experiment
\cite{SC-RTG}.
Additionally, in the experiment \cite{SC-RTG}, it has
been commented that the window of charge carrier density at which
superconductivity can be observed is narrower than the prediction.
We can understand the suppression directly from the transition temperature using Eq.~\eqref{eq:T_b} of the toy model, where the total number of charge carriers is halved within the energy cutoff $\epsilon_\mathrm{D}$, which is consistent with the experiment \cite{SC-RTG}. 
These pieces of evidence strongly support the notion that SC1 emerges at the band edge, which can be achieved through the opening of a gap by other order parameters, such as the IVC order parameters. 
This justifies that the IVC-SC state we have proposed is indeed a strong candidate for SC1 observed in RTG.

We also comment on possible implications when the chemical potential is within the gap. Near the phase boundary, when the mean-field temperature approaches zero, our toy model predicts that the conventional contribution remains finite while the quantum metric contribution diverges. 
{To verify our conclusions obtained using the toy model, we will conduct numerical calculations on the microscopic model, using Eqs.~(\ref{eq:TMF}-\ref{eq:l_c}).}

\subsection{Numerical calculations on the microscopic model} \label{sec:SC_IVC}
As a benchmark, we can calculate the property of the IVC-SC state in the microscopic model. We use a six-band microscopic model whose details are introduced in the supplementary material \cite{supp} and assume
that $|\boldsymbol{\Delta}_{\mathrm{IVC}}(\mathbf{q})|$ is independent of $\mathbf{q}$.
In general, one can solve for $|\boldsymbol{\Delta}_{\mathrm{IVC}}|$ using a self-consistency
equation, similar to the case of superconductivity. However, in our
calculation, we will assume that in the superconducting phase, $|\boldsymbol{\Delta}_{\mathrm{IVC}}|$
remains unchanged, and superconductivity occurs when in close vicinity
with the band edge of $\epsilon_{-}(\mathbf{q})$, thus a Cooper pair
is formed between $\psi_{-,\mathbf{q}}$ states \eqref{eq: IVC_State}. Using the fact that
at charge carrier density $n\approx-1.8\times10^{12}\mathrm{cm}^{-2}$ there is a sharp disappearance
of superconductivity, we can deduce that $|\boldsymbol{\Delta}_{\mathrm{IVC}}|\approx7$meV{, with details of fitting procedure discussed in the supplementary material \cite{supp}.}

With the $|\boldsymbol{\Delta}_{\mathrm{IVC}}|$ determined, we can calculate the
effective attractive interaction strength, using Eq.~(\ref{eq:TMF}),
with the dispersion set to $\epsilon_{-}(\mathbf{q})$. Numerically,
the mean-field temperature is the highest when the chemical potential
is around $22\mu\mathrm{eV}$ below the edge of the band. Setting the
maximal mean-field temperature as $125$mK, the effective attractive
interaction strength $U\approx1.2$eV, which could have originated from the fluctuation of the IVC order parameter \cite{IVC}. Similar to the flavour-symmetric state
calculation, with the interaction strength determined, using Eq.~\eqref{eq:TMF},
we can determine the mean-field temperature, as illustrated in Fig.~\ref{fig:3}(a).
The mean-field temperature calculated from the IVC state, in general,
agrees with the experimental result, both reaching a maximum at
$n\approx-1.85\times10^{12}\mathrm{cm}^{-2}$, and then decreasing down to $\approx70$mK
at $n\approx-1.93\times10^{12}\mathrm{cm}^{-2}$, and is no longer superconducting
for $n>-1.8\times10^{12}\mathrm{cm}^{-2}$. We note that for $n>-1.82\times10^{12}\mathrm{cm}^{-2}$,
where the chemical potential no longer cuts the IVC band that is responsible
for the formation of Cooper pair, superconductivity can still emerge from
finite temperature effect. We also note that the mismatch in mean-field temperature
could be due to contributions from other bands and a change in IVC order parameter as we tune the charge carrier density, which have been neglected in the numerical calculation.

We can calculate the coherence length using Eqs.~(\ref{eq:chi}-\ref{eq:l_c}),
thus checking if the corresponding upper critical field $H_{c2}$ matches with the experimental result. A plot of $H_{c2}$ at base temperature $50$mK, calculated
from the IVC state dispersion, is shown in Fig.~\ref{fig:3}(b).
We note that both the numerical calculations and
the experimental measurement of $H_{c2}$ have a maximal at density $n\approx-1.82\times10^{12}\mathrm{cm}^{-2}$.
Our calculation can also well capture the drop in $H_{c2}$ beyond
the band edge, with the little discrepancy possibly originating from the mismatch
in the mean-field temperature calculated numerically from the transition temperature
near $n\approx-1.82\times10^{12}\mathrm{cm}^{-2}$. 
We also comment
that the quantum metric could have a correctional effect on $H_{c2}$,
of approximately $5\%$ at $n\approx-1.82\times10^{12}\mathrm{cm}^{-2}$.
In fact, within the band gap, the variation in chemical potential is insignificant as illustrated in Fig~\ref{fig:2}(d), and the drop in $H_{c2}$ can mostly be accounted to the decrease in mean-field temperature. In particular, the experimental temperature $T=50$mK approaches the mean-field transition temperature at $n\approx-1.8\times10^{12}\mathrm{cm}^{-2}$, resulting in $H_{c2}=0$.
However, the quantum metric effect can set in if one considers the zero-temperature coherence length, that is, the Ginzburg-Landau coherence length $\xi_\mathrm{GL}$, which is calculated in Fig.~\ref{fig:3}(c). In most regions of charge carrier densities, the conventional contribution $\xi_\mathrm{GL,con}$ (blue line in Fig.~\ref{fig:3}) dominates and will approach a finite value. 
By comparison, the quantum metric contribution $\xi_\mathrm{GL,qm}$ will increase and finally diverge when the charge carrier density gets close to the phase boundary between the IVC-SC
state and IVC state.
This originates from the vanishingly low charge carrier density contributed by the relevant band at the Fermi energy when the chemical potential lies between the band gap and is very close to the band edge. 
Therefore, we can predict a divergent coherence length $\xi$ thus a vanishing upper critical field $H_{c2}$ when the lower temperature can be realized for experimental measurements around $n=-1.8\times 10^{12}$cm$^{12}$. {To conclude, the numerical results on the microscopic model agrees with qualitative insights previously obtained from the toy model.}

We should also remark that in our model, the IVC-SC state has effectively
extended the IVC order parameter into the flavour-symmetric phase, where IVC
state was originally not observed in higher temperature\cite{SC-RTG}. As such, contrary to the bare electron
picture where IVC and SC orders are competing orders, in the SC-IVC
state, where there exists an effective attractive interaction, thus Cooper
pair formation directly between the IVC states, the SC order combining
with the IVC order becomes more energy favorable than the flavor-symmetric phase.

\section{Discussion}
We propose a scenario of quasiparticle pairing of IVC for the SC1 of RTG. 
Our proposal can well resolve the discrepancies for the transition temperature and coherence length reported in experiments. 
We construct a toy model of a graphene-like system, with Dirac
points of Fermi velocity $v$ at the $K$ and $K'$ valleys, and
open up a mass gap $m$ at the Dirac points while assuming an s-wave pairing. 
We determined the transition temperature of s-wave superconductivity near the band gap with the behaviour
$T_{\mathrm{MF}}=\epsilon_D\exp(-2/\rho_\mathrm{qp}U)$ where $\rho_\mathrm{qp}$ is the density of states of IVC quasiparticles in comparison to the antiadiabatic limit of the BCS theory on the low-density charge carriers. 
Therefore, when the
chemical potential is near the mass gap, the transition temperature
is more suppressed, and the conventional coherence length is no longer
proportional to $v/T_{\mathrm{MF}}$, but instead $v/\sqrt{mT_{\mathrm{MF}}}$.
We also note that the quantum metric could have a comparable contribution
depending on the mean-field temperature, in comparison with the mass
gap.

{ We want to comment on the role of quantum metric, where the quantum metric only has a correction
effect in the current experimental setup, on the superconducting coherence length observed 
in RTG. However, near the phase transition from SC-IVC to IVC state,
the quantum metric contribution increases rapidly and diverges at
the phase boundary, while the conventional contribution remains finite. 
This serves as smoking-gun evidence for future experiments to be performed at a lower temperature, allowing distinction between conventional and quantum metric effect for $H_{c2}$.}

In a recent experiment \cite{BBG_Ising} of Bernal bilayer graphene with spin-orbit coupling, there is substantial evidence that superconductivity is prohibited when no neighbouring/coinciding PIP is observed, further supporting our claims that quasiparticles are essential for superconductivity near/within flavour-symmetry-breaking phases.
We believe our quasiparticle pairing picture can, in general, provide a phenomenological description for interaction-driven correlated phases, when there exists a neighbouring/overlapping flavour-symmetry-breaking phase, and, for example, possibly explain the discrepancy in the in-plane upper critical field between electron-doped and hole-doped superconductivity, observed in \cite{BBG_Ising}, which we leave to future work.

\section{Methods} \label{sec: Methodology}

\subsection{Mean-field theory of the Intervalley coherence state}
As discussed in Fig.~\ref{fig:1}(b), the density of states of electrons diverges
at the charge carrier density $n\approx-0.5\times10^{12}\mathrm{cm}^{-2}$. At this doping level,
the K and K$^\prime$ valley Fermi surfaces are almost perfectly nested together as illustrated in Fig.~\ref{fig:2}(c).
Assuming a repulsive interaction between K and K$^\prime$ valley, the intervalley
nesting would energetically favor the IVC state, for which a gap is
opened at the nested portion of the Fermi surfaces. 
To further study
the IVC state, we can define the mean-field Hamiltonian:
\begin{equation}
H_{\mathrm{MF}}=\sum_{\tau,\tau',\mathbf{q}}\psi_{\tau,\mathbf{q}}^{\dagger}h_{\tau,\tau'}(\mathbf{q})\psi_{\tau',\mathbf{q}},
\end{equation}
with the Hamiltonian matrix as
\begin{equation}
h_{\tau,\tau'}(\mathbf{q}) =\epsilon_{s}(\mathbf{q})I_{2}+\boldsymbol{\Delta}(\mathbf{q})\cdot\boldsymbol{\tau}.
\end{equation}
The $\boldsymbol{\Delta}(\mathbf q)$ is related to the IVC order parameter,
\begin{equation}
\!\!\!\!\boldsymbol{\Delta}(\mathbf{q})\!=(|\boldsymbol{\Delta}_{\mathrm{IVC}}(\mathbf{q})|\cos(\phi_{\mathbf{q}}),|\boldsymbol{\Delta}_{\mathrm{IVC}}(\mathbf{q})|\sin(\phi_{\mathbf{q}}),\epsilon_{a}(\mathbf{q})).
\label{eq:delta_q}
\end{equation}
 where $I_2$ is the $2\times2$ identity matrix and $\boldsymbol\tau=(\tau_1,\tau_2,\tau_3)$ are Pauli matrices. 
In Eq.~\eqref{eq:delta_q}, $\epsilon_{s}(\mathbf{q})=-\mu+\frac{1}{2}[\epsilon_{K}(\mathbf{q})+\epsilon_{K'}(\mathbf{q})]$
and $\epsilon_{a}(\mathbf{q})=\frac{1}{2}[\epsilon_{K}(\mathbf{q})-\epsilon_{K'}(\mathbf{q})]$
are the symmetric and antisymmetric parts of the flavor-symmetric state dispersion, respectively,
and $\Delta_{\mathrm{IVC}}(\mathbf{q})=|\boldsymbol{\Delta}_{\mathrm{IVC}}(\mathbf{q})|\exp(i\phi_{\mathbf{q}})$
is the order parameter for the IVC state. Note that the IVC order
parameter has a phase that can be well-matched with the phase of
the intervalley form factor \cite{IVC}.
The dispersion spectrum of the IVC state is given by:
\begin{equation}
    \epsilon_{\pm}(\mathbf{q})\approx \epsilon_{s}(\mathbf{q})\pm|\boldsymbol{\Delta}(\mathbf{q})|.
    \label{eq:qpdis}
\end{equation}

The dispersion spectrum for IVC states is shown in Fig.~\ref{fig:2}(d). The IVC states defined in the valley basis, correspond to two quasiparticle bands, with Bloch states:
\begin{align}
\psi_{+,\mathbf{q}} & =\frac{1}{\sqrt{2|\boldsymbol{\Delta}(\mathbf{q})|}}
\begin{pmatrix}
e^{-i\phi_{\mathbf{q}}}\sqrt{|\boldsymbol{\Delta}(\mathbf{q})|+\epsilon_{a}(\mathbf{q})}\\
\sqrt{|\boldsymbol{\Delta}(\mathbf{q})|-\epsilon_{a}(\mathbf{q})}
\end{pmatrix} , \\
\psi_{-,\mathbf{q}} & =\frac{1}{\sqrt{2|\boldsymbol{\Delta}(\mathbf{q})|}}\begin{pmatrix}-e^{-i\phi_{\mathbf{q}}}\sqrt{|\boldsymbol{\Delta}(\mathbf{q})|-\epsilon_{a}(\mathbf{q})}\\
\sqrt{|\boldsymbol{\Delta}(\mathbf{q})|+\epsilon_{a}(\mathbf{q})}
\end{pmatrix}.\label{eq: IVC_State}
\end{align}
Note that time reversal symmetry remains intact after the formation
of IVC state \cite{supp}, which is crucial for the existence of
superconductivity in our proposal by assuming an s-wave pairing. We should also retrieve an effective attractive interaction, to enable the formation of Cooper pairs. 
By adopting the argument in Ref.~\onlinecite{RG-IVC}, involving the introduction of a weaker anti-ferromagnetic Hund's coupling compared to the repulsive intervalley interaction, it becomes possible to achieve an effective attractive intervalley interaction. This ultimately results in the emergence of a spin-singlet superconducting phase. We comment that the interaction between quasiparticle could also have originated from the fluctuations of the quasiparticle order parameter, as illustrated in Ref.~\onlinecite{IVC}.
As shown in Fig.~\ref{fig:2}(d), the density of states of the quasiparticles peak at the edge of the valence band (the blue line), which allows us to introduce a toy model of massive Dirac cones with the chemical potential close to the band edge.

\subsection{Ginzburg-Landau theory on toy models}
We mimic the IVC-SC state with a simplified toy model. 
In our toy model, similar to RTG, we have two Dirac
cones, of Fermi velocity $v$, centered at the $K$ and $K^\prime$
valley respectively. By acting with an external potential, for example
a displacement field, we open up a mass gap $m$ between the originally
The system can be described effectively in the
continuum limit by introducing an IVC order to massive Dirac fermion Hamiltonian at $K$ and $K'$ valleys. At mean-field level,  the Hamiltonian $H_\mathrm{IVC}=\int \frac{d^2 q}{(2\pi)^2}\Psi_s^\dagger(\mathbf q) \mathcal H_s(\mathbf q) \Psi_s(\mathbf q)$
for spin $s=\uparrow,\;\downarrow$ is given by assuming degeneracy for the two spin sectors:
\begin{align}
    \mathcal H_s &= 
    \begin{pmatrix}
        h_+(\mathbf{q})
        &
        \Delta(\mathbf{q})
        \\
        \Delta^\dagger(\mathbf{q})
        &
         h_-(\mathbf{q})
    \end{pmatrix}
\end{align}
where $ \Delta(\mathbf{q}) = \Delta_{\mathrm{IVC}}\;\mathrm{diag}([e^{-i\phi_q},-e^{-i\phi_q}])$ with IVC order parameter $\Delta_{\mathrm{IVC}}$ and $qe^{i\phi}=q_x+i q_y$. We use the valley-sublattice basis $\Psi_s= [a_{+sA},a_{+sB},a_{-sA},a_{-sB}]^T$ with $h_\tau(\mathbf{q})$ being the massive Dirac fermion Hamiltonian, of valley indices $\tau=\pm$, and $\mathbf{q}$ the momentum difference to the center of the valley. In specific, $h_\tau(\mathbf{q})$ is given by:
\begin{align}
h_{\tau}(\mathbf{q}) & =v\mathbf{q}\cdot(\tau \sigma_{x}\hat{\mathbf{e}}_{x}+\sigma_{y}\hat{\mathbf{e}}_{y})+m\sigma_{z},
\end{align}
where $\sigma_{x,y,z}$ correspond to the Pauli matrix defined in the orbital basis, and $\bar\tau$ refers to the opposite valley. 
In particular, we will focus on the quasiparticle band of dispersion spectrum $E(\mathbf q)=\sqrt{v^{2}q^{2}+m^{2}}\pm\Delta_{\mathrm{IVC}}$, assuming $\Delta_{\mathrm{IVC}}\ll m$ and
consider only on-site effective attractive interaction between time-reversal copy of quasiparticles. This can be realized as intervalley interaction for a graphene heterostructure:
\begin{equation}
H_{\mathrm{int}}=-U\sum_{s\lambda\lambda^\prime}\int d^{2}\mathbf{r}\ a_{+s\lambda}^{\dagger}(\mathbf{r})a_{-\bar s\lambda^\prime}^{\dagger}(\mathbf{r})a_{-\bar s\lambda^\prime }(\mathbf{r})a_{+s\lambda}(\mathbf{r}),
\label{eq:Hint}
\end{equation}
where $U$ is attractive interaction strength, and $a_{\tau s \lambda}(\mathbf{r})$
is the quasiparticle annihilation operator of the valley $\tau$, spin $s$ and sublattice index $\lambda$. Due to the large band gap separating the conduction and valence band, we will focus on the contributions of the valence bands of the toy model,
which can be realized via a projection by involving the Bloch waves $|u_{\tau}(\mathbf q)\rangle$ of the valence bands \cite{GL,ACL_QM},
\begin{equation}
a_{\tau s\lambda}(\mathbf{r})\rightarrow\int\frac{d^{2}\mathbf{q}}{(2\pi)^{2}}e^{i\mathbf{q}\cdot\mathbf{r}}u_{\tau \lambda}^{*}(\mathbf{q})c_{ s}(\mathbf{q}),
\label{eq:projection}
\end{equation}
where $c_{s}(\mathbf{q})$ is the quasiparticle annihilation operator
of momentum $\mathbf{q}$ for the valence band. The time reversal symmetry is intact for an IVC quasiparticle band, $u_{\tau \lambda}(\mathbf q)=u^*_{\bar\tau\lambda}(-\mathbf q)$, and thus the effective attractive interaction is coupling between IVC quasiparticles of opposite momentum.
Via the Hubbard-Stratonovich
transformation~\cite{SM-TG}, we can then introduce a bosonic field $\Delta(\mathbf{r})= \sum_s c_{\uparrow}(\mathbf{r})c_{ \downarrow}(\mathbf{r})$
that corresponds to Cooper pairs for the projected Hamiltonian.  
Using the path integral formalism,
we can evaluate the free energy of the system,
which can be decomposed into two components: the mean-field solution and the fluctuations. 
This decomposition is achieved by expanding the free energy around its extremum. Further details of the calculation can be found in the Supplementary Information \cite{supp}.
Focusing on the case around the transition region where the mean-field value of the bosonic field
$\Delta_{0}$ vanishes, minimizing the mean-field
solution of the free energy provides us with a self-consistency condition
for the mean-field temperature, at which transition
to the superconducting state initiates:
\begin{equation}
1=U\int\frac{d^{2}\mathbf{q}}{(2\pi)^{2}}\frac{1}{2\varepsilon(\mathbf{q})}\tanh\frac{\beta_{\mathrm{MF}}\varepsilon(\mathbf{q})}{2},
\label{eq:TMF}
\end{equation}
where $T_{\mathrm{MF}}=1/\beta_{\mathrm{MF}}$ is the mean-field temperature,
and for convenience we have defined $\varepsilon(\mathbf{q})=\epsilon_{0}(\mathbf{q})-\mu$.
The fluctuation of the free energy, in particular the quadratic term
$F_{2}$, can be related to the fluctuation of the order parameter
$\delta\Delta(\mathbf{k})$. After integrating out the fermionic fields
$F_{2}$ takes the form:
\begin{equation}
 F_{2} =\int\frac{d^{2}\mathbf{k}}{(2\pi)^{2}}|\delta\Delta(\mathbf{k})|^{2}\left[U-U^{2}\chi(\mathbf{k})\right],
\label{eq:F2}    
\end{equation}
where $\chi(\mathbf{k})$ is the four-point coherence function,
\begin{align}
\chi(\mathbf{k}) & =\int\frac{d^{2}\mathbf{q}}{(2\pi)^{2}}|\Gamma(\mathbf{q},\mathbf{k})|^{2}\nonumber\\
&\phantom{=}\times\frac{\tanh\left[\frac{\beta}{2}\varepsilon\left(\mathbf{q}+\frac{\mathbf{k}}{2}\right)\right]+\tanh\left[\frac{\beta}{2}\varepsilon\left(\mathbf{q}-\frac{\mathbf{k}}{2}\right)\right]}{2\left[\varepsilon\left(\mathbf{q}+\frac{\mathbf{k}}{2}\right)+\varepsilon\left(\mathbf{q}-\frac{\mathbf{k}}{2}\right)\right]}.
\end{align}
And we have defined the form factor $|\Gamma(\mathbf{q},\mathbf{k})|^{2}=\left|u_{c,+}^{T}\left(\mathbf{q}+\frac{\mathbf{k}}{2}\right)u_{c,-}\left(-\mathbf{q}+\frac{\mathbf{k}}{2}\right)\right|^{2}$,
which was introduced upon performing projection \eqref{eq:projection}
from a multiband system onto a single relevant band. Given time reversal
invariant of the Bloch states $u^*_{c,-}(-\mathbf{q})=u_{c,+}(\mathbf{q})\equiv u(\mathbf q)$,
we can expand the form factor around $\mathbf{k}=0$ as \cite{GL}:
\begin{equation}
|\Gamma(\mathbf{q},\mathbf{k})|^{2}=1-\sum_{i,j}k_{i}k_{j}g_{ij}(\mathbf{q}),
\end{equation}
where $g_{ij}(\mathbf{q})=\frac{1}{2}\mathrm{Tr}\left[\partial_{i}P(\mathbf{q})\partial_{j}P(\mathbf{q})\right]$
is the quantum metric, and $P(\mathbf{q})=\left|u(\mathbf{q})\right\rangle\left\langle u(\mathbf{q})\right|$ is the projection matrix
for the Bloch state. Similarly, we can expand the four-point coherence
function $\chi(\mathbf{k})$ around $\mathbf{k}=0$, with the assumption
that the valence band is rotational invariant as: 
\begin{equation}
    \chi(\mathbf{k}) =\chi_{0}-\left(\chi_{2,\mathrm{con}}+\chi_{2,\mathrm{qm}}\right)k^{2}+\mathcal{O}(k^{4}),
    \label{eq:chi}
\end{equation}
where $\chi_{2,\mathrm{con}}$ is the contribution due to dispersion
of the band, and $\chi_{2,\mathrm{qm}}$ is due to the form factor from the quantum metric. Recalling the definition of $F_{2}$ (\ref{eq:F2}),
we can construct an effective Lagrangian that governs the fluctuation
of the order parameter $\delta\Delta(\mathbf{k})$ formed as $F_{2}=\int\frac{d^{2}\mathbf{k}}{(2\pi)^{2}}\mathcal{L}[\delta\Delta]$:
\begin{align}
\mathcal{L}[\delta\Delta] & =\frac{1}{2m^{*}}|\nabla\delta\Delta|^{2}+\alpha|\delta\Delta|^{2}+\mathcal{O}(|\delta\Delta|^{4}),
\end{align}
with the coefficients
\begin{align}
\frac{1}{2m^{*}} & =U^{2}\left(\chi_{2,\mathrm{con}}+\chi_{2,\mathrm{qm}}\right),\\
\alpha & =U-U^{2}\chi_{0}.
\end{align}
From the effective Lagrangian, we can determine the general form of
the superconducting coherence length, as motivated by the Ginzburg-Landau
theory:
\begin{align}
\xi & =\sqrt{\frac{U^{2}\left(\chi_{2,\mathrm{con}}+\chi_{2,\mathrm{qm}}\right)}{|U-U^{2}\chi_{0}|}}=\sqrt{\xi_{\mathrm{con}}^{2}+\xi_{\mathrm{qm}}^{2}},
\label{eq:l_c}
\end{align}
where the overall coherence length is governed by both the conventional
contribution and the quantum metric~\cite{GL,ACL_QM}. 
The conventional contribution $\xi_\mathrm{con}$ depends on the band dispersion with a similar form as a conventional s-wave superconductor, and it vanishes in the flat-band limit. The quantum metric part $\xi_\mathrm{qm}$ exclusively appears as a multi-band effect, and it is determined by the quantum metric $g_{ij}$.
The coherence length $\xi$  will be scaled as $\xi=(\frac{T_{\mathrm{MF}}-T}{T_{\mathrm{MF}}})^{-1/2}\xi_\mathrm{GL}$ since the Ginzburg-Landau theory is developed around $T\rightarrow T_\mathrm{MF}$, while we extract the Ginzburg-Landau coherence length $\xi_\mathrm{GL}$ as in the standard way.  
We comment that Eqs.~(\ref{eq:projection}-\ref{eq:l_c}) are applicable to any microscopic model, such as an IVC state in RTG, which hosts two valleys with dispersion $\epsilon_+(\mathbf{q})=\epsilon_-(\mathbf{-q})$, and has local attractive interaction between time-reversal copies of the same quasiparticle, which can be realized with onsite intervalley interaction given by Eq.~\eqref{eq:Hint}.
In the Sec.~Result, we have studied the case that is relevant to the IVC-SC state, for which
the mass gap $m\gg T_{\mathrm{MF}}$, in particular when the chemical potential is near the band gap. Details of the analytical calculation and
a brief discussion of the $m\ll T_{\mathrm{MF}}$ regime are included in
the supplementary material \cite{supp}.
\\
Below we will illustrate with the case of $m\gg T_{\mathrm{MF}}$, when the chemical potential satisfies $\beta_{\mathrm{MF}}(\mu-m)\gg1$,
thus away from the mass gap. The mean-field temperature can be solved
self-consistently by Eq.~\eqref{eq:TMF}:
\begin{equation}
T_{\mathrm{MF}} \propto\sqrt{\rho_\mathrm{qp}\epsilon_{D}}\exp\left(-\frac{1}{\rho_\mathrm{qp} U'}\right),
\end{equation}
Note that as $\mu\rightarrow\epsilon_{D}$, the BCS
relationship for mean-field temperature $T_\mathrm{MF}\propto\epsilon_{D}\exp\left(-1/\rho_\mathrm{qp} U\right)$ can be recovered.
We can also derive the conventional coherence length $\xi_{\mathrm{con}}$,
which is proportional to $v/T_{\mathrm{MF}}$, similar to the BCS
limit: 
\begin{equation}
\xi_{\mathrm{con}}\approx\frac{v}{T_{\mathrm{MF}}}\left(\frac{T_{\mathrm{MF}}-T}{T_{\mathrm{MF}}}\right)^{-1/2}\sqrt{\frac{7\zeta(3)}{32\pi^{2}}},
\end{equation}
where $\zeta$ is the Riemann zeta function and the $\xi_\mathrm{GL,con}\propto \frac{v}{T_\mathrm{MF}}$. As for the quantum metric contribution, it can be approximated by:
\begin{equation}
\xi_{\mathrm{qm}}\approx\frac{v}{4\mu}\left(\frac{T_{\mathrm{MF}}-T}{T_{\mathrm{MF}}}\right)^{-1/2}\sqrt{\ln\left(\frac{\mu^{3}}{mT_{\mathrm{MF}}^{2}}\right)},
\end{equation}
where the factor $\sqrt{\ln({\mu^{3}}/{mT_{\mathrm{MF}}^{2}})}$ is in order of unity. As such, the conventional contribution dominates, by a factor of $\sim\beta_{\mathrm{MF}}\mu\gg1$.
Thus, the quantum metric effect is negligible, and the system can be
treated as conventional when away from the band edge.

\section*{ Data Availability}
This is a theoretical study and no experimental datasets were generated or analysed. All results can be reproduced from the equations and parameters provided in the main text and Supplementary Information.

\nocite{*}
\bibliography{draft}

\section*{Acknowledgements}
C.W.C.~acknowledges funding from the Croucher Cambridge International Scholarship by the Croucher Foundation and the Cambridge Trust. K. T. L. acknowledges the support of the Ministry of Science and Technology, China, The New Cornerstone Foundation, and the Hong Kong Research Grants Council through Grants No. MOST23SC01-A, No. RFS2021-6S03, No. C6053-23G, No. AoE/P-701/20, AoE/P-604/25R, No. 16309223, No. 16311424 and No. 16300325.

\section*{Author Contributions Statement}
S.C.\ and K.T.L.\ proposed and supervised the project. C.W.C.\ conducted the calculations and analysis. S.C.\ and C.W.C.\ wrote the manuscript.

\section*{Competing Interests Statement}
The authors declare no competing interests.

\newpage
\onecolumngrid
\newpage
\appendix
{
\begin{center}
\textbf{Supplementary Information of ``Superconductivity from Quasiparticle Pairing of Intervalley Coherent State in Rhombohedral Trilayer Graphene"}   
\end{center} 
}

\section{Details of 6-band Hamiltonian for RTG}

Following the definition of the sub-lattice A and B in the single layer
of graphene, we can denote the structure of RTG by stating which pairs
of sites coincide from the top view:
\begin{equation}
    B_{1}\leftrightarrow A_{2},\ B_{2}\leftrightarrow A_{3},\ B_{3}\leftrightarrow A_{1}
\end{equation}
As such we can classify our hopping \cite{BS} ($S_{i}\in\{A_{i},B_{i}\}$):
\begin{align*}
\gamma_{0}: & A_{i}\leftrightarrow B_{i} ,\quad   i=1,2,3\\
\gamma_{1}: & B_{i}\leftrightarrow A_{i+1} ,\quad    i=1,2\\
\gamma_{2}: & A_{1}\leftrightarrow B_{3}\\
\gamma_{3}: & A_{i}\leftrightarrow B_{i+1} ,\quad   i=1,2\\
\gamma_{4}: & S_{i}\leftrightarrow S_{i+1} ,\quad    i=1,2\\
\gamma_{5}: & B_{1}\leftrightarrow A_{3}\\
\gamma_{6}: & S_{1}\leftrightarrow S_{3}
\end{align*}
Due to direct interlayer hopping $\gamma_{1}$ between $B_{i}$ and
$A_{i+1}$, sites not involved, namely $A_{1}$ and $B_{3}$, would
have lower energy. We denote the energy difference of $A_{1}(B_{3})$
from $B_{1}(A_{3})$ as $\delta<0$. Note that these hopping are listed
in decreasing magnitude, with $\gamma_{3}$ and $\gamma_{4}$ having
similar order. $\gamma_{5}$ and $\gamma_{6}$ having similar order
as well, but should be much weaker in comparison. 

We should also note the massless Dirac fermion contribution near the
valleys, for a single layer graphene takes the form:
\begin{equation}
H=v_{0}\begin{pmatrix}0 & \pi^{\dagger}\\
\pi & 0
\end{pmatrix},
\end{equation}
where we have defined $\pi=\tau p_{x}+ip_{y}$, with $\tau=+(-)$ correspond
to the $K(K')$ valley. With the single layer picture and list of
possible hopping in mind, we can construct our basis as $(A_{1},B_{3},B_{1},A_{2},B_{2},A_{3})$
for convenience:
\begin{equation}
\mathcal{H}=\begin{pmatrix}u_{1}+\delta & \gamma_{2}/2 & v_{0}\pi^{\dagger} & v_{4}\pi^{\dagger} & v_{3}\pi & v_{6}\pi\\
\gamma_{2}/2 & u_{3}+\delta & v_{6}\pi^{\dagger} & v_{3}\pi^{\dagger} & v_{4}\pi & v_{0}\pi\\
v_{0}\pi & v_{6}\pi & u_{1} & \gamma_{1} & v_{4}\pi^{\dagger} & v_{5}\pi^{\dagger}\\
v_{4}\pi & v_{3}\pi & \gamma_{1} & u_{2} & v_{0}\pi^{\dagger} & v_{4}\pi^{\dagger}\\
v_{3}\pi^{\dagger} & v_{4}\pi^{\dagger} & v_{4}\pi & v_{0}\pi & u_{2} & \gamma_{1}\\
v_{6}\pi^{\dagger} & v_{0}\pi^{\dagger} & v_{5}\pi & v_{4}\pi & \gamma_{1} & u_{3}
\end{pmatrix},
\end{equation}
where $v_{i}=\sqrt{3}a\gamma_{i}/2\hbar$, as motivated by single
layer graphene. ($a=0.246\mathrm{nm}$)

Note that for convenience, we will take $v_{i}=\sqrt{3}a\gamma_{i}/2$.
And define $\mathbf{q}$ as the wavevector instead of using momentum
$\mathbf{p}$ to compensate for the factor of $\hbar$. 
The low-energy effective Hamiltonian is:
\begin{align}
H_{\mathrm{eff}} & =H_{\mathrm{ch}}+H_{\mathrm{s}}+H_{\mathrm{tr}}+H_{\mathrm{gap}}+H_{\mathrm{s}}^{\prime},\\
H_{\mathrm{ch}} & =\frac{(v_{0}q)^{3}}{\gamma_{1}^{2}}[\cos(3\varphi_{\mathbf{q}})\sigma_{x}+\sin(3\varphi_{\mathbf{q}})\sigma_{y}],\\
H_{\mathrm{s}} & =\left(\delta-\frac{2v_{0}v_{4}q^{2}}{\gamma_{1}}\right)\sigma_{0},\\
H_{\mathrm{tr}} & =\left(\frac{\gamma_{2}}{2}-\frac{2v_{0}v_{3}q^{2}}{\gamma_{1}}\right)\sigma_{x},\\
H_{\mathrm{gap}} & =u_{d}\left[1-\left(\frac{v_{0}q}{\gamma_{1}}\right)^{2}\right]\sigma_{z},\\
H_{\mathrm{s}}^{\prime} & =\frac{u_{a}}{3}\left[1-3\left(\frac{v_{0}q}{\gamma_{1}}\right)^{2}\right]\sigma_{0},
\end{align}
where we have defined $\tan\varphi_{\mathbf{q}}=q_{y}/q_{x}$, $u_{d}=(u_{1}-u_{3})/2$
and $u_{a}=(u_{1}+u_{3})/2-u_{2}$. In our scope of discussion, the
contribution from $\gamma_{5}$ and $\gamma_{6}$ are assumed to be
small thus neglected.

We will use parameters fitted in \cite{HM-RTG}, with the parameters 
listed in Table~\ref{tab-1}.
The parameter $u_{d}$, which is the energy difference between each layer, can be
induced via an external displacement field \cite{ECP-RTG}: $u_{d}=\frac{ez}{\epsilon_{r}}D $,
where $z=0.33$nm is the interlayer distance, and $\epsilon_{r}=4.4$
is the dielectric constant hexagonal boron nitride. In our study,
we will focus on superconductivity of the valence band when $D=0.46$V/nm,
which corresponds to $u_{d}=34.5$meV. 
\begin{table}[t]
\begin{centering}
\caption{{Tight-binding Parameters for Rhombohedral Trilayer
Graphene. Retrieved from \cite{HM-RTG}.}}
\begin{tabular}{|c|c|}
\hline 
Hopping parameters & Values (eV)\tabularnewline
\hline 
$\delta$ & -0.00105\tabularnewline
\hline 
$\gamma_{0}$ & 3.1\tabularnewline
\hline 
$\gamma_{1}$ & 0.38\tabularnewline
\hline 
$\gamma_{2}$ & -0.015\tabularnewline
\hline 
$\gamma_{3}$ & -0.29\tabularnewline
\hline 
$\gamma_{4}$ & -0.141\tabularnewline
\hline 
$u_{a}$ & -0.0023\tabularnewline
\hline 
\end{tabular}\label{tab-1}
\par\end{centering}
\end{table}

\section{Derivation of free energy}
We mimic the IVC-SC state by a simplified toy model. 
In our toy model, similar to RTG, we have two Dirac
cones, of Fermi velocity $v$, centered at the $K$ and $K^\prime$
valley respectively. By acting with an external potential, for example
a displacement field, we open up a mass gap $m$ between the originally
touching Dirac bands. 
The system can be described effectively in the
continuum limit by introducing intervalley density-density interaction to massive Dirac fermion Hamiltonian at $K$ and $K'$ valleys, which at mean-field level for spin $s=\uparrow,\;\downarrow$ is given by:
\begin{align}
    \mathcal H_s &= 
    \begin{pmatrix}
        h_+(\mathbf{q})
        &
        \Delta(\mathbf{q})
        \\
        \Delta^*(\mathbf{q})
        &
         h_-(\mathbf{q})
    \end{pmatrix}
    \nonumber\\
    \Delta(\mathbf{q})
    &=
    \sum_n
    \Delta_{\mathrm{IVC}}\tilde{u}_{n,-}(\mathbf{q})\tilde{u}^\dagger_{n,+}(\mathbf{q})
\end{align}
where we have assumed degeneracy for the two spin sectors, with IVC order parameter $\Delta$. Here we use the valley-sublattice basis $\Psi_s= [a_{+sA},a_{+sB},a_{-sA},a_{-sB}]^T$. With $h_\tau(\mathbf{q})$ being the massive Dirac fermion Hamiltonian, of valley indices $\tau=\pm$, $\mathbf{q}$ the momentum difference to the center of the valley, and $\tilde{u}_{n,\tau}(\mathbf{q})$ being the Bloch vector of Hamiltonian $h_\tau(\mathbf{q})$ of band index $n$. In specific, $h_\tau(\mathbf{q})$ is given by:
\begin{align}
h_{\tau}(\mathbf{q}) & =v\mathbf{q}\cdot(\tau\sigma_{x}\hat{\mathbf{e}}_{x}+\sigma_{y}\hat{\mathbf{e}}_{y})+m\sigma_{z},
\end{align}
where $v$ is the fermi velocity before gap-opening, $\tau=+/-$
correspond to $K$ and $K'$ valley respectively, $m$ is the gap
size, and $\mathbf{q}$ is the momentum difference with respect to
the center of the valley, and $\sigma_{i}$ Pauli matrices that act
on the sublattice basis. Corresponding Bloch states in sublattice
basis of $\mathcal{H}_s(\mathbf{q})$ are given by:
\begin{align}
u_{c, \eta,\tau}(\mathbf{q}) & =\frac{1}{\sqrt{2\epsilon_{0}(\mathbf{q})}}
\begin{pmatrix}
e^{-i\tau\phi/2}\sqrt{\epsilon_{0}(\mathbf{q})+m}\\
\tau e^{i\tau\phi/2}\sqrt{\epsilon_{0}(\mathbf{q})-m}
\end{pmatrix}  
\otimes
\frac{1}{\sqrt{2}}
\begin{pmatrix}
    \eta \\ 1
\end{pmatrix}
&E&=  \epsilon_{0}(\mathbf{q}) + \eta\Delta_{\mathrm{IVC}},\\
u_{v,\eta,\tau}(\mathbf{q}) & =\frac{1}{\sqrt{2\epsilon_{0}(\mathbf{q})}}
\begin{pmatrix}-e^{-i\tau\phi/2}\sqrt{\epsilon_{0}(\mathbf{q})-m}\\
\tau e^{i\tau\phi/2}\sqrt{\epsilon_{0}(\mathbf{q})+m}
\end{pmatrix}   
\otimes
\frac{1}{\sqrt{2}}
\begin{pmatrix}
    \eta \\ 1
\end{pmatrix}
&E&=  -\epsilon_{0}(\mathbf{q}) + \eta\Delta_{\mathrm{IVC}},
\end{align}
where $\epsilon_{0}(\mathbf{q})$ is given by $\sqrt{v^{2}q^{2}+m^{2}}$,
index $c$ ($v$) denotes the conduction (valence) band, $\eta=\pm$ denote the pair of quasiparticle bands, and have
rewritten our momentum in polar coordinate as $qe^{i\phi_q}=q_{x}+iq_{y}$. Note the explicit assumption that the Hamiltonian is the same for spin up and down fermions. We can also write down the exact form of the IVC order parameter:
\begin{equation}
    \Delta(\mathbf{q})
    =
    \Delta_{\mathrm{IVC}}
    \begin{pmatrix}
        e^{i\phi_q}   &   0   \\
        0           &   -e^{-i\phi_q}
    \end{pmatrix}
\end{equation}
To induce superconductivity in the system, we introduce an effective
attractive interaction between quasiparticles:
\begin{equation}
H_{\mathrm{int}}=-U\sum_{s\lambda\lambda^\prime}\int d^{2}\mathbf{r}\ a_{+s\lambda}^{\dagger}(\mathbf{r})a_{-\bar s\lambda^\prime}^{\dagger}(\mathbf{r})a_{-\bar s\lambda^\prime }(\mathbf{r})a_{+s\lambda}(\mathbf{r}),\label{Hint}
\end{equation}
where $a_{\tau s \lambda}(\mathbf{r})$ is the fermionic annihilation
operator of states in valley of momentum $\tau\mathbf{K}$ of spin $s=\uparrow,\;\downarrow$ and sublattice $\lambda = A,~B$
at position $\mathbf{r}$. Below to simplify the calculation we choose $s=\uparrow$, which has no effect in the overall result, and we will not write down the spin index explicitly.
As such we can write down a partition function
involving only fermionic fields $a$: 
\begin{align}
Z & =\int D[a,\bar{a}]e^{-\int_{0}^{\beta}d\tau\int d\mathbf{r}\mathcal{L}_{a}[a,\bar{a}]},\\
\mathcal{L}_{a}[a,\bar{a}] & =\mathcal{L}_{\mathrm{Dirac}}[a,\bar{a}]+(\partial_{\tau}-\mu)[\bar{a}_{-}(\mathbf{r})a_{-}(\mathbf{r})+\bar{a}_{+}(\mathbf{r})a_{+}(\mathbf{r})] \notag\\
 & -U\bar{a}_{-}(\mathbf{r})\bar{a}_{+}(\mathbf{r})a_{+}(\mathbf{r})a_{-}(\mathbf{r}),
\end{align}
where $\bar{a}$ and $a$ are Grassmann field. Using the Hubbard-Stratonovich
(HS) transformation, we can introduce a Bosonic field: 
\begin{equation}
1=\int D[\Delta,\bar{\Delta}]e^{-U\int_{0}^{\beta}d\tau\int d\mathbf{r}[\bar{\Delta}(\mathbf{r})-\bar{a}_{-}(\mathbf{r})\bar{a}_{+}(\mathbf{r})][\Delta(\mathbf{r})-a_{+}(\mathbf{r})a_{-}(\mathbf{r})]}.
\end{equation}
As such we can rewrite our partition function as: 
\begin{align}
Z & =\int D[\Delta,\bar{\Delta}]e^{-U\int_{0}^{\beta}d\tau\int d\mathbf{r}|\boldsymbol{\Delta}(\mathbf{r})|^{2}}\mathcal{Z}[\Delta,\bar{\Delta}],\\
\mathcal{Z}[\Delta,\bar{\Delta}] & =\int D[a,\bar{a}]e^{-\int_{0}^{\beta}d\tau\int d\mathbf{r}\mathcal{L}[\Delta,\bar{\Delta},a,\bar{a}]},\\
\mathcal{L}[\Delta,\bar{\Delta},a,\bar{a}] & =\mathcal{L}_{\mathrm{Dirac}}[a,\bar{a}]+(\partial_{\tau}-\mu)[\bar{a}_{-}(\mathbf{r})a_{-}(\mathbf{r})+\bar{a}_{+}(\mathbf{r})a_{+}(\mathbf{r})]\nonumber \\
 & \ -U[\bar{\Delta}(\mathbf{r})a_{+}(\mathbf{r})a_{-}(\mathbf{r})+\mathrm{h.c.}].
\end{align}
We can introduce the projection of an annihilation operator onto the
relevant band as: 
\begin{equation}
a_{\tau s\lambda}(\mathbf{r})\rightarrow\mathcal{A}\int\frac{d^{2}\mathbf{q}}{(2\pi)^{2}}e^{i\mathbf{q}\cdot\mathbf{r}}u_{v,-,\tau}^{*}(\mathbf{q})c(\mathbf{q}),
\end{equation}
where $c_{\tau}(\mathbf{q})$ is the the band electron annihilation
operator of momentum $\mathbf{q}$, and $\mathcal{A}$ is the unit
cell area of the lattice. As such we can project our Lagrangian density
to $\mathcal{L}[\Delta,\bar{\Delta},c,\bar{c}]$: 
\begin{align}
\mathcal{L}[\Delta,\bar{\Delta},a,\bar{a}]  \rightarrow&\mathcal{L}[\Delta,\bar{\Delta},c,\bar{c}]\nonumber \\
 =& [\epsilon_{0}(\mathbf{q})+\partial_{\tau}-\mu][\bar{c}(\mathbf{q})c(\mathbf{q})+\bar{c}(\mathbf{q})c(\mathbf{q})] \nonumber \\
 &-U\mathcal{A}\int\frac{d^{2}\mathbf{k}}{(2\pi)^{2}}\left[\bar{\Delta}(\mathbf{k})\Gamma(\mathbf{q},\mathbf{k})c\left(\mathbf{q}+\frac{\mathbf{k}}{2}\right)c\left(-\mathbf{q}+\frac{\mathbf{k}}{2}\right)+\mathrm{h.c.}\right],
\end{align}
where we have redefined $c_{\tau}(\mathbf{q})$ as the corresponding
Grassmann field. Note that the form factor is defined by $\Gamma(\mathbf{q},\mathbf{k})=u_{v,-,+}^{T}\left(\mathbf{q}+\frac{\mathbf{k}}{2}\right)u_{v,-,-}\left(-\mathbf{q}+\frac{\mathbf{k}}{2}\right)$.
We have also assume the bosonic field $\Delta(\mathbf{q)}$ transforms
trivially as a spin 0 boson $\Delta(\mathbf{r)}\rightarrow\mathcal{A}\int\frac{d^{2}\mathbf{k}}{(2\pi)^{2}}e^{i\mathbf{k}\cdot\mathbf{r}}\Delta(\mathbf{k)}$.
We can rewrite our Lagrangian density into a matrix form using the
Nambu spinors:
\begin{equation}
\mathcal{L}[\Delta,\bar{\Delta},c,\bar{c}]=\mathcal{A}\int\frac{d^{2}\mathbf{q}'}{(2\pi)^{2}}\begin{pmatrix}\bar{c}(\mathbf{q')} & c(\mathbf{-q')}\end{pmatrix}G_{\mathbf{q}',\mathbf{q}}\begin{pmatrix}c(\mathbf{q})\\
\bar{c}(\mathbf{-q})
\end{pmatrix}.
\end{equation}
Defining the free energy of the system as $Z=\int D[\Delta,\bar{\Delta}]e^{-\beta F[\Delta,\bar{\Delta}]}$,
or equivalently as:
\begin{align}
F[\Delta,\bar{\Delta}] & =\int\frac{d^{2}\mathbf{k}}{(2\pi)^{2}}U|\boldsymbol{\Delta}(\mathbf{k})|^{2}-T\ln\int D[c,\bar{c}]\ e^{-\int_{0}^{\beta}d\tau\int\frac{d^{2}\mathbf{q}}{(2\pi)^{2}}\mathcal{L}[\Delta,\bar{\Delta},c,\bar{c}]}\nonumber \\
 & =\int\frac{d^{2}\mathbf{k}}{(2\pi)^{2}}U|{\Delta}(\mathbf{k})|^{2}-T\int\frac{d^{2}\mathbf{q}}{(2\pi)^{2}}\int\frac{d^{2}\mathbf{q}'}{(2\pi)^{2}}\ln\det G_{\mathbf{q},\mathbf{q}'}.
\end{align}
To calculate the free energy, we expand the Bosonic field $\Delta(\mathbf{k})$
around the extremum of the free energy as:
\begin{equation}
\Delta(\mathbf{k})=\Delta_{0}\delta(\mathbf{k})+\delta\Delta(\mathbf{k}),
\end{equation}
where $\Delta_{0}$ is the mean-field solution, real when properly
gauged, and $\delta\Delta(\mathbf{k})$ is the fluctuation. We can
thus decompose our Lagrangian density $\mathcal{L}[\Delta,\bar{\Delta},c,\bar{c}]$
as:
\begin{align}
\mathcal{L}[\Delta,\bar{\Delta},c,\bar{c}] & =\mathcal{L}_{0}+\delta\mathcal{L}\\
\mathcal{L}_{0} & =[\epsilon_{0}(\mathbf{q})+\partial_{\tau}-\mu][\bar{c}(\mathbf{q})c(\mathbf{q})+\bar{c}(\mathbf{q})c(\mathbf{q})]-U[\Gamma(\mathbf{q},0)\Delta_{0}c(\mathbf{q})c(-\mathbf{q})+\mathrm{h.c.}]\\
\delta\mathcal{L} & =-U\int\frac{d^{2}\mathbf{k}}{(2\pi)^{2}}\left[\Gamma(\mathbf{q},\mathbf{k})\delta\bar{\Delta}(\mathbf{k})c\left(\mathbf{q}+\frac{\mathbf{k}}{2}\right)c\left(-\mathbf{q}+\frac{\mathbf{k}}{2}\right)+\mathrm{h.c.}\right].
\end{align}
Correspondingly, we can decompose the free energy $F[\Delta,\bar{\Delta}]$
as:
\begin{equation}
F[\Delta,\bar{\Delta}]=F_{0}+\delta F,
\end{equation}
where $F_{0}$ would be given by:
\begin{align}
F_{0} & =\int\frac{d^{2}\mathbf{q}}{(2\pi)^{2}}\left\{ U|\boldsymbol{\Delta}_{0}|^{2}-T\sum_{n}\ln\left[\beta^{2}\det\begin{pmatrix}\epsilon_{0}(\mathbf{q})-i\omega_{n}-\mu & -U\Gamma(\mathbf{q},0)\Delta_{0}\\
-(U\Gamma(\mathbf{q},0)\Delta_{0})^{*} & -\epsilon_{0}(\mathbf{q})-i\omega_{n}+\mu
\end{pmatrix}\right]\right\} \nonumber \\
 & =\int\frac{d^{2}\mathbf{q}}{(2\pi)^{2}}\left\{ U|\Delta_{0}|^{2}-T\sum_{n}\ln\beta^{2}[-\omega_{n}^{2}-(\epsilon_{0}(\mathbf{q})-\mu)^{2}-|U\Gamma(\mathbf{q},0)\Delta_{0}|^{2}]\right\} \nonumber \\
 & =\int\frac{d^{2}\mathbf{q}}{(2\pi)^{2}}\left\{ U|\Delta_{0}|^{2}-T\sum_{n}\left\{ \ln\beta[-i\omega_{n}+\epsilon(\mathbf{q})]+\ln\beta[-i\omega_{n}-\epsilon(\mathbf{q})]\right\} \right\} \nonumber \\
 & =\int\frac{d^{2}\mathbf{q}}{(2\pi)^{2}}\left\{ U|\Delta_{0}|^{2}-T\left[\ln\left(1+e^{-\beta\epsilon(\mathbf{q})}\right)+\ln\left(1+e^{\beta\epsilon(\mathbf{q})}\right)\right]\right\} \nonumber \\
 & =\int\frac{d^{2}\mathbf{q}}{(2\pi)^{2}}\left[U|\Delta_{0}|^{2}-\epsilon(\mathbf{q})-2T\ln\left(1+e^{-\beta\epsilon(\mathbf{q})}\right)\right].
\end{align}
Here $\epsilon(\mathbf{q})=\sqrt{(\epsilon_{0}(\mathbf{q})-\mu)^{2}+|U\Gamma(\mathbf{q},0)\Delta_{0}|^{2}}$
and $\omega_{n}=2n\pi T$ is the Matsubara frequencies. Note that
$F_{0}$ recovers the form of the grand potential. We should also
mention the self-consistency condition for $\Delta_{0}$, namely that
$\frac{\partial F_{0}}{\partial\Delta_{0}}=0$ gives us:
\begin{align}
0 & =\int\frac{d^{2}\mathbf{q}}{(2\pi)^{2}}\left[2U\Delta_{0}-\frac{\partial\epsilon(\mathbf{q})}{\partial\Delta_{0}}\left(1-\frac{2e^{-\beta\epsilon(\mathbf{q})}}{1+e^{-\beta\epsilon(\mathbf{q})}}\right)\right]\nonumber \\
 & =\int\frac{d^{2}\mathbf{q}}{(2\pi)^{2}}\left[2U\Delta_{0}-\frac{U^{2}|\Gamma(\mathbf{q},0)|^{2}\Delta_{0}}{\epsilon(\mathbf{q})}\tanh\frac{\beta\epsilon(\mathbf{q})}{2}\right]\nonumber \\
 & =\int\frac{d^{2}\mathbf{q}}{(2\pi)^{2}}\left[1-\frac{U|\Gamma(\mathbf{q},0)|^{2}}{2\epsilon(\mathbf{q})}\tanh\frac{\beta\epsilon(\mathbf{q})}{2}\right]\nonumber \\
1 & =\mathcal{\mathcal{A}}\int\frac{d^{2}\mathbf{q}}{(2\pi)^{2}}\frac{U|\Gamma(\mathbf{q},0)|^{2}}{2\epsilon(\mathbf{q})}\tanh\frac{\beta\epsilon(\mathbf{q})}{2}.\label{SC}
\end{align}
The higher order contribution $\delta F$ is given by:
\begin{align}
\delta F & =\int\frac{d^{2}\mathbf{k}}{(2\pi)^{2}}\left[U|\delta\Delta|^{2}-T\ln\frac{\int D[c,\bar{c}]\ e^{-\int_{0}^{\beta}d\tau\int\frac{d^{2}\mathbf{q}}{(2\pi)^{2}}(\mathcal{L}_{0}+\delta\mathcal{L})}}{\int D[c,\bar{c}]\ e^{-\int_{0}^{\beta}d\tau\int\frac{d^{2}\mathbf{q}}{(2\pi)^{2}}\mathcal{L}_{0}}}\right]\nonumber \\
 & =\int\frac{d^{2}\mathbf{k}}{(2\pi)^{2}}\left[U|\delta\Delta|^{2}-T\left\langle e^{-\int_{0}^{\beta}d\tau\int\frac{d^{2}\mathbf{q}}{(2\pi)^{2}}\delta\mathcal{L}}-1\right\rangle \right],
\end{align}
where we have defined the thermal average $\langle f\rangle=\left(\int D[c,\bar{c}]\ e^{-\int_{0}^{\beta}d\tau\int\frac{d^{2}\mathbf{q}}{(2\pi)^{2}}\mathcal{L}_{0}}\right)^{-1}\int D[c,\bar{c}]\ e^{-\int_{0}^{\beta}d\tau\int\frac{d^{2}\mathbf{q}}{(2\pi)^{2}}\mathcal{L}_{0}}f$.
To facilitate the calculation of Gor'kov's Green function thus $\delta F$,
we should introduce generating functional:
\begin{equation}
\mathcal{Z}[\eta,\bar{\eta}]=\int D[\psi,\bar{\psi}]\ e^{-\int_{0}^{\beta}d\tau\int\frac{d^{2}\mathbf{q}}{(2\pi)^{2}}\left[\bar{\psi}(\mathbf{q})G_{0}(\mathbf{q})\psi(\mathbf{q})-\bar{\eta}(\mathbf{q})\psi(\mathbf{q})-\bar{\psi}(\mathbf{q})\eta(\mathbf{q})\right]},
\end{equation}
where $\bar{\psi}(\mathbf{q)}=\begin{pmatrix}\bar{c}_{-}(\mathbf{q}) & c_{+}(\mathbf{-q})\end{pmatrix}$
is the Nambu spinor, and $\mathcal{L}_{0}=\bar{\psi}(\mathbf{q})G_{0}(\mathbf{q})\psi(\mathbf{q})$.
By performing shifting on the field $\psi(\mathbf{q})\rightarrow\psi(\mathbf{q})+G_{0}^{-1}(\mathbf{q})\eta(\mathbf{q})$
and $\bar{\psi}(\mathbf{q})\rightarrow\bar{\psi}(\mathbf{q})+\bar{\eta}(\mathbf{q})G_{0}^{-1}(\mathbf{q})$,
we arrive at:
\begin{equation}
\mathcal{Z}[\eta,\bar{\eta}]=\mathcal{Z}[0,0]\ e^{\int_{0}^{\beta}d\tau\int\frac{d^{2}\mathbf{q}}{(2\pi)^{2}}\bar{\eta}(\mathbf{q})G_{0}^{-1}(\mathbf{q})\eta(\mathbf{q})}.
\end{equation}
As such thermal average $\langle\psi_{\mathbf{q}}\bar{\psi}_{\mathbf{q}'}\rangle$
takes the form:
\begin{align}
\langle\psi_{\mathbf{q}}\bar{\psi}_{\mathbf{q}'}\rangle & =\left.\frac{\delta}{\delta\eta(\mathbf{q}')}\frac{\delta}{\delta\bar{\eta}(\mathbf{q})}e^{\int_{0}^{\beta}d\tau\int\frac{d^{2}\mathbf{q}}{(2\pi)^{2}}\bar{\eta}(\mathbf{k})G_{0}^{-1}(\mathbf{k})\eta(\mathbf{k})}\right|_{\eta,\bar{\eta}=0}\nonumber \\
 & =\left.\frac{\delta}{\delta\eta(\mathbf{q}')}\left[G_{0}^{-1}(\mathbf{q})\eta(\mathbf{q})e^{\int_{0}^{\beta}d\tau\int\frac{d^{2}\mathbf{q}}{(2\pi)^{2}}\bar{\eta}(\mathbf{k})G_{0}^{-1}(\mathbf{k})\eta(\mathbf{k})}\right]\right|_{\eta,\bar{\eta}=0}\nonumber \\
 & =G_{0}^{-1}(\mathbf{q})\delta_{\mathbf{q},\mathbf{q'}}.
\end{align}
where $G_{0}^{-1}(\mathbf{q})$ has the expression:
\begin{equation}
G_{0}^{-1}(\mathbf{q})=\frac{1}{\omega_{n}^{2}+\epsilon^{2}(\mathbf{q})}\begin{pmatrix}\epsilon_{0}(\mathbf{q})+i\omega_{n}-\mu & -U\Gamma(\mathbf{q},0)\Delta_{0}\\
-U\Gamma^{*}(\mathbf{q},0)\Delta_{0} & -\epsilon_{0}(\mathbf{q})+i\omega_{n}+\mu
\end{pmatrix}.
\end{equation}
As such, we have Gor'kov's Green functions:
\begin{align}
\mathcal{G}(\mathbf{q}) & =\langle c(\mathbf{q})\bar{c}(\mathbf{q})\rangle
=\frac{i\omega_{n}+\epsilon_{0}(\mathbf{q})-\mu}{\omega_{n}^{2}+\epsilon^{2}(\mathbf{q})}, \label{G}\\
\mathcal{F}(\mathbf{q}) & =\langle c(-\mathbf{q})c(\mathbf{q})\rangle=-\frac{U\Gamma^{*}(\mathbf{q},0)\Delta_{0}}{\omega_{n}^{2}+\epsilon^{2}(\mathbf{q})}.\label{F}
\end{align}
Note that due to time reversal invariant, $\Gamma(\mathbf{q},0)=1$,
thus $\epsilon(\mathbf{q})=\sqrt{(\epsilon_{0}(\mathbf{q})-\mu)^{2}+U^{2}\Delta_{0}^{2}}$.
With all information, we can solve for $\delta F$. Due to stability
of the mean-field solution, a linear correction has to vanish, and the
lowest order term is thus second order in $\delta\Delta(\mathbf{k})$,
namely $F_{2}$ the Gaussian fluctuation of the free energy, given
by:
\begin{align}
F_{2} & =\int\frac{d^{2}\mathbf{k}}{(2\pi)^{2}}U|\delta\Delta(\mathbf{k})|^{2}-\frac{T}{2}\left\langle \left(\int_{0}^{\beta}d\tau\int\frac{d^{2}\mathbf{k}}{(2\pi)^{2}}\delta\mathcal{L}\right)^{2}\right\rangle \nonumber \\
 & =\int\frac{d^{2}\mathbf{k}}{(2\pi)^{2}}U|\delta\Delta(\mathbf{k})|^{2}\nonumber \\
 & \ -\frac{T}{2}\sum_{n}\mathcal{A}\int\frac{d^{2}\mathbf{q}}{(2\pi)^{2}}\int\frac{d^{2}\mathbf{k}}{(2\pi)^{2}}\left[2|\Gamma(\mathbf{q},\mathbf{k})|^{2}|\delta\Delta(\mathbf{k})|^{2}\left\langle c\left(\mathbf{q}+\frac{\mathbf{k}}{2}\right)\bar{c}\left(\mathbf{q}+\frac{\mathbf{k}}{2}\right)\right\rangle \left\langle c\left(\mathbf{-q}+\frac{\mathbf{k}}{2}\right)\bar{c}\left(\mathbf{-q}+\frac{\mathbf{k}}{2}\right)\right\rangle \right.\nonumber \\
 & \ +\Gamma(\mathbf{q},\mathbf{k})\Gamma(\mathbf{q},-\mathbf{k})\delta\Delta(\mathbf{k})\delta\Delta(-\mathbf{k})\left\langle \bar{c}\left(\mathbf{q}+\frac{\mathbf{k}}{2}\right)\bar{c}\left(\mathbf{-q}-\frac{\mathbf{k}}{2}\right)\right\rangle \left\langle \bar{c}\left(\mathbf{-q}+\frac{\mathbf{k}}{2}\right)\bar{c}\left(\mathbf{q}-\frac{\mathbf{k}}{2}\right)\right\rangle \nonumber \\
 & \ \left.+\Gamma^{*}(\mathbf{q},\mathbf{k})\Gamma^{*}(\mathbf{q},-\mathbf{k})\delta\Delta^{*}(\mathbf{k})\delta\Delta^{*}(-\mathbf{k})\left\langle c\left(-\mathbf{q}+\frac{\mathbf{k}}{2}\right)c\left(\mathbf{q}+\frac{\mathbf{k}}{2}\right)\right\rangle \left\langle c\left(\mathbf{q}+\frac{\mathbf{k}}{2}\right)c\left(-\mathbf{q}-\frac{\mathbf{k}}{2}\right)\right\rangle \right]\nonumber \\
 & =\int\frac{d^{2}\mathbf{k}}{(2\pi)^{2}}|\delta\Delta(\mathbf{k})|^{2}\left[U-U^{2}\chi(\mathbf{k})\right]\label{F2}\\
\chi(\mathbf{q}) & =T\sum_{n}\mathcal{A}\int\frac{d^{2}\mathbf{q}}{(2\pi)^{2}}|\Gamma(\mathbf{q},\mathbf{k})|^{2}\left[\mathcal{G}\left(\mathbf{q}+\frac{\mathbf{k}}{2}\right)\mathcal{G}\left(-\mathbf{q}+\frac{\mathbf{k}}{2}\right)+\mathcal{F}\left(\mathbf{q}+\frac{\mathbf{k}}{2}\right)\mathcal{F}\left(-\mathbf{q}+\frac{\mathbf{k}}{2}\right)\right]\label{chi}.
\end{align}

\section{Quantum metric of massive Dirac cones}

Assuming time reversal invariant, form factor $|\Gamma(\mathbf{q},\mathbf{k})|^{2}$
can be expanded given that $\mathbf{k}$ is small as:
\begin{align}\label{qm}
|\Gamma(\mathbf{q},\mathbf{k})|^{2} & =\left|\left\langle u_{\mathbf{q}+\mathbf{k}/2}\right|\left.u_{\mathbf{q}-\mathbf{k}/2}\right\rangle \right|^{2}\nonumber \\
 & =\mathrm{Tr}\left[P\left(\mathbf{q}+\frac{\mathbf{k}}{2}\right)P\left(\mathbf{q}-\frac{\mathbf{k}}{2}\right)\right]\nonumber \\
 & =\mathrm{Tr}\left\{ \left[P(\mathbf{q})+\sum_{i}\frac{k_{i}}{2}\partial_{i}P(\mathbf{q})+\sum_{i,j}\frac{k_{i}k_{j}}{8}\partial_{i}\partial_{j}P(\mathbf{q})\right]\left[P(\mathbf{q})-\sum_{a}\frac{k_{a}}{2}\partial_{a}P(\mathbf{q})+\sum_{a,b}\frac{k_{a}k_{b}}{8}\partial_{a}\partial_{b}P(\mathbf{q})\right]\right\} \nonumber \\
 & =1-\sum_{i,j}\frac{k_{i}k_{j}}{8}\mathrm{Tr}\left[2\partial_{i}P(\mathbf{q})\partial_{j}P(\mathbf{q})-P(\mathbf{q})\partial_{i}\partial_{j}P(\mathbf{q})-\partial_{i}\partial_{j}P(\mathbf{q})P(\mathbf{q})\right]\nonumber \\
 & =1-\sum_{i,j}\frac{k_{i}k_{j}}{8}\mathrm{Tr}\left[4\partial_{i}P(\mathbf{q})\partial_{j}P(\mathbf{q})-\partial_{i}\partial_{j}P(\mathbf{q})\right]\nonumber \\
 & =1-\sum_{i,j}\frac{1}{2}k_{i}k_{j}\mathrm{Tr}\left[\partial_{i}P(\mathbf{q})\partial_{j}P(\mathbf{q})\right]+\frac{k_{i}k_{j}}{8}\partial_{i}\partial_{j}\mathrm{Tr}P(\mathbf{q})\nonumber \\
  & =1-\sum_{i,j}\frac{1}{2}k_{i}k_{j}\mathrm{Tr}\left[\partial_{i}P(\mathbf{q})\partial_{j}P(\mathbf{q})\right]\nonumber \\
 & =1-\sum_{i,j}k_{i}k_{j}g_{ij}(\mathbf{q}),
\end{align}
where $g_{ij}(\mathbf{q})=\frac{1}{2}\mathrm{Tr}\left[\partial_{i}P(\mathbf{q})\partial_{j}P(\mathbf{q})\right]$
is the quantum metric. For our toy model, in terms of elementary matrices, the projector matrix
is given by:
\begin{align}
P & =\frac{1}{2\epsilon_{0}(\mathbf{q})}\begin{pmatrix}e^{-i\phi/2}\sqrt{\epsilon_{0}(\mathbf{q})+m}\\
e^{i\phi/2}\sqrt{\epsilon_{0}(\mathbf{q})-m}
\end{pmatrix}\begin{pmatrix}e^{i\phi/2}\sqrt{\epsilon_{0}(\mathbf{q})+m} & e^{-i\phi/2}\sqrt{\epsilon_{0}(\mathbf{q})-m}\end{pmatrix}\nonumber \\
 & =\frac{1}{2\epsilon_{0}(\mathbf{q})}\begin{pmatrix}\epsilon_{0}(\mathbf{q})+m & e^{-i\phi}\sqrt{\epsilon_{0}(\mathbf{q})^{2}-m^{2}}\\
e^{i\phi}\sqrt{\epsilon_{0}(\mathbf{q})^{2}-m^{2}} & \epsilon_{0}(\mathbf{q})-m
\end{pmatrix}\nonumber \\
 & =\frac{1}{2\epsilon_{0}(\mathbf{q})}\begin{pmatrix}\epsilon_{0}(\mathbf{q})+m & vqe^{-i\phi}\\
vqe^{i\phi} & \epsilon_{0}(\mathbf{q})-m
\end{pmatrix}\nonumber \\
&=\frac{1}{2}\left(\mathds{1}_{2}+\frac{m}{\epsilon_{0}(\mathbf{q})}\sigma_{z}+\frac{v}{\epsilon_{0}(\mathbf{q})}\mathbf{q}\cdot\boldsymbol{\sigma}\right).
\end{align}
As such the derivative is given by:
\begin{align}
\partial_{i}P(\mathbf{q}) & =\frac{1}{2\epsilon_{0}(\mathbf{q})}\left[v\sigma_{i}-\frac{\partial_{i}\epsilon_{0}(\mathbf{q})}{\epsilon_{0}(\mathbf{q})}(m\sigma_{z}+v\mathbf{q}\cdot\boldsymbol{\sigma})\right]\nonumber \\
 & =\frac{1}{2\epsilon_{0}(\mathbf{q})}\left[v\sigma_{i}-\frac{v^{2}q_{i}}{\epsilon_{0}(\mathbf{q})^{2}}(m\sigma_{z}+v\mathbf{q}\cdot\boldsymbol{\sigma})\right]\nonumber \\
 & =\frac{v}{2\epsilon_{0}(\mathbf{q})^{3}}\left[\epsilon_{0}(\mathbf{q})^{2}\sigma_{i}-vq_{i}(m\sigma_{z}+v\mathbf{q}\cdot\boldsymbol{\sigma})\right]\nonumber \\
 & =\frac{v}{2\epsilon_{0}(\mathbf{q})^{3}}\left[\left(m^{2}+v^{2}q_{j}^{2}+iv^{2}q_{i}q_{j}\sigma_{z}\epsilon_{ij}\right)\sigma_{i}-mvq_{i}\sigma_{z}\right]\\
 & =\frac{v}{2\epsilon_{0}(\mathbf{q})^{3}}\left[\sigma_{i}\left(m^{2}+v^{2}q_{j}^{2}-iv^{2}q_{i}q_{j}\sigma_{z}\epsilon_{ij}\right)-mvq_{i}\sigma_{z}\right],
\end{align}
where $j\neq i$, and $\epsilon_{xy}=-\epsilon_{yx}=1$. Correspondingly
quantum geometry is:
\begin{align}
g_{ii}(\mathbf{q}) & =\frac{1}{2}\mathrm{Tr}[\partial_{i}P(\mathbf{q})\partial_{i}P(\mathbf{q})]\nonumber \\
 & =\frac{v^{2}}{8\epsilon_{0}(\mathbf{q})^{6}}\left[\left(m^{2}+v^{2}q_{j}^{2}\right)^{2}+v^{4}q_{i}^{2}q_{j}^{2}+m^{2}v^{2}q_{i}^{2}\right]\mathrm{Tr}\mathds{1}_{2}\nonumber \\
 & =\frac{v^{2}}{4\epsilon_{0}(\mathbf{q})^{6}}\left(m^{4}+2m^{2}v^{2}q_{j}^{2}+v^{4}q_{j}^{4}+v^{4}q_{i}^{2}q_{j}^{2}+m^{2}v^{2}q_{i}^{2}\right)\nonumber \\
 & =\frac{v^{2}}{4\epsilon_{0}(\mathbf{q})^{6}}\left(m^{4}+m^{2}v^{2}q_{j}^{2}+v^{4}q^{2}q_{j}^{2}+m^{2}v^{2}q^{2}\right)\nonumber \\
 & =\frac{v^{2}}{4\epsilon_{0}(\mathbf{q})^{4}}\left(m^{2}+v^{2}q_{j}^{2}\right)\\
g_{ij}(\mathbf{q},i\neq j) & =\frac{1}{2}\mathrm{Tr}[\partial_{i}P(\mathbf{q})\partial_{i}P(\mathbf{q})]\nonumber \\
 & =\frac{v^{2}}{8\epsilon_{0}(\mathbf{q})^{6}}\left\{ \mathrm{Tr}\left[i\epsilon_{ij}\sigma_{z}\left(m^{2}+v^{2}q_{i}^{2}+iv^{2}q_{i}q_{j}\sigma_{z}\epsilon_{ij}\right)\left(m^{2}+v^{2}q_{j}^{2}+iv^{2}q_{i}q_{j}\sigma_{z}\epsilon_{ij}\right)\right]+2m^{2}v^{2}q_{i}q_{j}\right\} \nonumber \\
 & =\frac{v^{2}}{8\epsilon_{0}(\mathbf{q})^{6}}\left\{ \mathrm{Tr}\left[i\epsilon_{ij}\sigma_{z}\left(2m^{2}+v^{2}q^{2}\right)iv^{2}q_{i}q_{j}\sigma_{z}\epsilon_{ij}\right]+2m^{2}v^{2}q_{i}q_{j}\right\} \nonumber \\
 & =\frac{v^{2}}{4\epsilon_{0}(\mathbf{q})^{6}}\left[-\left(2m^{2}+v^{2}q^{2}\right)v^{2}q_{i}q_{j}+m^{2}v^{2}q_{i}q_{j}\right]\nonumber \\
 & =\frac{v^{2}}{4\epsilon_{0}(\mathbf{q})^{6}}\left[-\left(2m^{2}+v^{2}q^{2}\right)v^{2}q_{i}q_{j}+m^{2}v^{2}q_{i}q_{j}\right]\nonumber \\
 & =-\frac{v^{2}}{4\epsilon_{0}(\mathbf{q})^{4}}v^{2}q_{i}q_{j}\\
g(\mathbf{q}) & =\frac{v^{2}}{4(m^{2}+v^{2}q^{2})^{2}}\begin{pmatrix}m^{2}+v^{2}q_{y}^{2} & -v^{2}q_{x}q_{y}\\
-v^{2}q_{x}q_{y} & m^{2}+v^{2}q_{x}^{2}
\end{pmatrix}.
\end{align}

We will use this result to compare between the conventional and the
geometrical contribution in the later session. 

\section{Ginzburg-Landau Theory of Massive Dirac Cone}

At the upper critical field $H_{c2}$, the order parameter is suppressed,
namely $\Delta_{0}=0$. In this situation, $\epsilon(\mathbf{q})=|\epsilon_{0}(\mathbf{q})-\mu|$,
and Gor'kov's anomalous Green Function \eqref{F} vanishes, with the normal Green Function \eqref{G}
simplifies to:
\begin{equation}
\mathcal{G}(\mathbf{q})=\frac{1}{-i\omega_{n}+\epsilon_{0}(\mathbf{q})-\mu}.
\end{equation}
Recall Eq.~\eqref{F2} where the free energy density takes the
form $F_{2}=\int\frac{d^{2}\mathbf{k}}{(2\pi)^{2}}|\delta\Delta(\mathbf{k})|^{2}\left[U-U^{2}\chi(\mathbf{k})\right]$,
where:
\begin{align}
\chi(\mathbf{k}) & =T\sum_{n}\mathcal{A}\int\frac{d^{2}\mathbf{q}}{(2\pi)^{2}}|\Gamma(\mathbf{q},\mathbf{k})|^{2}\mathcal{G}\left(\mathbf{q}+\frac{\mathbf{k}}{2}\right)\mathcal{G}\left(-\mathbf{q}+\frac{\mathbf{k}}{2}\right)\nonumber \\
 & =T\mathcal{A}\int\frac{d^{2}\mathbf{q}}{(2\pi)^{2}}|\Gamma(\mathbf{q},\mathbf{k})|^{2}\sum_{n}\frac{1}{-i\omega_{n}+\epsilon_{0}\left(\mathbf{q}+\frac{\mathbf{k}}{2}\right)-\mu}\frac{1}{i\omega_{n}+\epsilon_{0}\left(\mathbf{-q}+\frac{\mathbf{k}}{2}\right)-\mu}\nonumber \\
 & =\frac{1}{2}\mathcal{A}\int\frac{d^{2}\mathbf{q}}{(2\pi)^{2}}|\Gamma(\mathbf{q},\mathbf{k})|^{2}\frac{\tanh\frac{\beta}{2}\left[\epsilon_{0}\left(\mathbf{q}+\frac{\mathbf{k}}{2}\right)-\mu\right]+\tanh\frac{\beta}{2}\left[\epsilon_{0}\left(\mathbf{q}-\frac{\mathbf{k}}{2}\right)-\mu\right]}{\epsilon_{0}\left(\mathbf{q}+\frac{\mathbf{k}}{2}\right)+\epsilon_{0}\left(\mathbf{q}-\frac{\mathbf{k}}{2}\right)-2\mu}\nonumber \\
 & =\mathcal{A}\int\frac{d^{2}\mathbf{q}}{(2\pi)^{2}}|\Gamma(\mathbf{q},\mathbf{k})|^{2}\frac{\tanh\left[\frac{\beta}{2}\varepsilon\left(\mathbf{q}+\frac{\mathbf{k}}{2}\right)\right]+\tanh\left[\frac{\beta}{2}\varepsilon\left(\mathbf{q}-\frac{\mathbf{k}}{2}\right)\right]}{2\left[\varepsilon\left(\mathbf{q}+\frac{\mathbf{k}}{2}\right)+\varepsilon\left(\mathbf{q}-\frac{\mathbf{k}}{2}\right)\right]}.
\end{align}
In the last line, we have defined $\varepsilon(\mathbf{q})=\epsilon_{0}(\mathbf{q})-\mu$
for convenience. Note that the conduction band dispersion is invariant
under rotational symmetry, thus isotropic. As such, we can perform
expansion around $\mathbf{k}=0$ to obtain:
\begin{align}
\chi(\mathbf{k}) & =\chi_{0}-\left(\chi_{2,\mathrm{con}}+\chi_{2,\mathrm{qm}}\right)k^{2}+\mathcal{O}(k^{4})\label{supp_eq:chi},\\
\chi_{0} & =\mathcal{\mathcal{A}}\int\frac{d^{2}\mathbf{q}}{(2\pi)^{2}}\frac{1}{2\varepsilon(\mathbf{q})}\tanh\frac{\beta\varepsilon(\mathbf{q})}{2},\\
\chi_{2,\mathrm{qm}}k^{2} & =\frac{k^{2}}{2}\mathcal{A}\int\frac{d^{2}\mathbf{q}}{(2\pi)^{2}}\mathrm{tr}g(\mathbf{q})\frac{\tanh\left[\frac{\beta}{2}\varepsilon\left(\mathbf{q}\right)\right]}{2\varepsilon\left(\mathbf{q}\right)},\\
\chi_{2,\mathrm{con}}k^{2} & =\mathcal{A}\int\frac{d^{2}\mathbf{q}}{(2\pi)^{2}}\left\{ \frac{\tanh\left[\frac{\beta}{2}\varepsilon\left(\mathbf{q}\right)\right]}{2\varepsilon\mathbf{(q})}-\frac{\tanh\left[\frac{\beta}{2}\varepsilon\left(\mathbf{q}+\frac{\mathbf{k}}{2}\right)\right]+\tanh\left[\frac{\beta}{2}\varepsilon\left(\mathbf{q}-\frac{\mathbf{k}}{2}\right)\right]}{2\left[\varepsilon\left(\mathbf{q}+\frac{\mathbf{k}}{2}\right)+\varepsilon\left(\mathbf{q}-\frac{\mathbf{k}}{2}\right)\right]}\right\} ,
\end{align}
where $\chi_{2,\mathrm{con}}$ is contribution due to dispersion of
the band, and $\chi_{2,\mathrm{qm}}$ is due to the form factor, thus
the quantum metric. Note the zeroth order term can be rewritten as:
\begin{align}
\chi_{0} & =\mathcal{A}\int\frac{d^{2}\mathbf{q}}{(2\pi)^{2}}\frac{\tanh\left[\frac{\beta}{2}\varepsilon\left(\mathbf{q}\right)\right]}{2\varepsilon\left(\mathbf{q}\right)}\nonumber \\
 & =\mathcal{A}\int\frac{d^{2}\mathbf{q}}{(2\pi)^{2}}\frac{\tanh\left[\frac{\beta}{2}\varepsilon\left(\mathbf{q}\right)\right]-\tanh\left[\frac{\beta_{\mathrm{MF}}}{2}\varepsilon\left(\mathbf{q}\right)\right]}{2\varepsilon\left(\mathbf{q}\right)}+\frac{1}{U},
\end{align}
where $T_{\mathrm{MF}}=\frac{1}{\beta_{\mathrm{MF}}}$is the mean-field
critical temperature, defined by the continuous version of the self-consistency
condition \eqref{SC} when $\Delta_{0}=0$:
\begin{align}
\frac{1}{U} & =\mathcal{A}\int\frac{d^{2}\mathbf{q}}{(2\pi)^{2}}\frac{1}{2\varepsilon(\mathbf{q})}\tanh\left[\frac{\beta_{\mathrm{MF}}}{2}\varepsilon(\mathbf{q})\right].
\end{align}
In the following, we will perform analytical calculations for the nearly
massless regime, namely the mass gap of the Dirac bands, is much smaller
than the Debye energy, which defines the energy range where we have
effective attractive interaction. In particular, we will study the
case where the mass gap $m\gg T_{\mathrm{MF}}$ and $m\ll T_{\mathrm{MF}}$.

\subsection{Case 1: $m\gg T_\mathrm{MF}$}

In continuous model, we can treat the chemical potential as a free
parameter. In particular, we will set it to $|\mu|=m+\kappa\epsilon_{D}$,
and with $0\leq\kappa\leq1$. To begin with we should write down our
integral over momentum space in terms of energy:
\begin{align}
\int\frac{d^{2}\mathbf{q}}{(2\pi)^{2}} =\int_{m}^{m+(1+\nu)\epsilon_{D}}\frac{\epsilon_{0}d\epsilon_{0}}{2\pi v^{2}}
 =\int_{-\kappa\epsilon_{D}}^{\epsilon_{D}}\frac{(\varepsilon+\mu)d\varepsilon}{2\pi v^{2}}.
\end{align}
Considering the contribution from quantum metric, we have:
\begin{align}
\chi_{2,\mathrm{qm}}k^{2} & =\frac{k^{2}}{2}\mathcal{A}\int\frac{d^{2}\mathbf{q}}{(2\pi)^{2}}(g_{xx}+g_{yy})\frac{\tanh\left[\frac{\beta}{2}\varepsilon(\mathbf{q})\right]}{2\varepsilon(\mathbf{q})}\nonumber \\
 & =\frac{1}{2}k^{2}\mathcal{A}\int\frac{d^{2}\mathbf{q}}{(2\pi)^{2}}\frac{v^{2}(2m^{2}+v^{2}q^{2})}{4(m^{2}+v^{2}q^{2})^{2}}\frac{\tanh\left[\frac{\beta}{2}\varepsilon(\mathbf{q})\right]}{2\varepsilon(\mathbf{q})}\nonumber \\
 & =\frac{1}{4}k^{2}\mathcal{A}\int_{-\kappa\epsilon_{D}}^{\epsilon_{D}}\frac{d\varepsilon}{2\pi}\frac{1}{2}\frac{m^{2}+\left(\varepsilon+\mu\right)^{2}}{\left(\varepsilon+\mu\right)^{3}}\frac{\tanh\left({\normalcolor \frac{\beta}{2}\varepsilon}\right)}{2\varepsilon}\nonumber \\
 & =\frac{1}{4}k^{2}\mathcal{A}\left[\int_{-\kappa\epsilon_{D}}^{\epsilon_{D}}\frac{d\varepsilon}{2\pi}\frac{1}{2}\frac{m^{2}}{\left(\varepsilon+\mu\right)^{3}}\frac{\tanh\left({\normalcolor \frac{\beta}{2}\varepsilon}\right)}{2\varepsilon}+\int_{-\kappa\epsilon_{D}}^{\epsilon_{D}}\frac{d\varepsilon}{2\pi}\frac{1}{2}\frac{\tanh\left({\normalcolor \frac{\beta}{2}\varepsilon}\right)}{2\varepsilon\left(\varepsilon+\mu\right)}\right].
\end{align}

For $\kappa\beta\epsilon_{D}\gg1$, the latter term dominates:
\begin{align}
\chi_{2,\mathrm{qm}}k^{2} & =\frac{1}{4}k^{2}\mathcal{A}\int_{-\kappa\epsilon_{D}}^{\epsilon_{D}}\frac{d\varepsilon}{2\pi}\frac{1}{2}\frac{\tanh\left({\normalcolor \frac{\beta}{2}\varepsilon}\right)}{2\varepsilon\left(\varepsilon+\mu\right)}\nonumber \\
 & =\frac{1}{32\pi}k^{2}\mathcal{A}\int_{-\kappa\epsilon_{D}}^{\epsilon_{D}}d\varepsilon\ \frac{1}{\mu}\left(\frac{1}{\varepsilon}-\frac{1}{\varepsilon+\mu}\right)\tanh\left(\frac{\beta}{2}\varepsilon\right)\nonumber \\
 & \approx\frac{1}{32\pi}k^{2}\mathcal{A}\int_{-\kappa\epsilon_{D}}^{\epsilon_{D}}d\varepsilon\ \frac{1}{\mu}\frac{1}{\varepsilon}\tanh\left(\frac{\beta}{2}\varepsilon\right)-\frac{1}{16\pi}k^{2}\mathcal{A}\int_{-\kappa\epsilon_{D}}^{\epsilon_{D}}d\varepsilon\ \frac{1}{\mu}\frac{1}{\varepsilon+\mu}\nonumber \\
 & \approx\frac{1}{32\pi}k^{2}\mathcal{A}\frac{1}{\mu}\left[2\ln\left(\frac{4e^{\gamma}}{\pi}\frac{\beta\epsilon_{D}}{2}\right)+\ln\kappa-\ln\frac{(\epsilon_{D}+\mu)m}{\mu^{2}}\right]\nonumber \\
 & \approx\frac{k^{2}\mathcal{A}}{32\pi\mu}\left[\ln\left(\frac{\beta^{2}\mu^{3}\epsilon_{D}}{m(\mu+\epsilon_{D})}\right)+2\ln\left(\frac{2e^{\gamma}}{\pi}\right)\right]\nonumber\\
 & \approx\frac{k^2\mathcal{A}}{32\pi\mu}\ln\left(\frac{\beta^{2}\mu^{3}}{m}\right).
\end{align}
where we have used the approximation $\int_{0}^{a}dx\tanh(x)/x\approx\ln\left(4e^{\gamma}a/\pi\right)$
for $a\gg1$. 

For $\kappa\beta\epsilon_{D}\ll1$, both term contributes. We can
appoximate the values of each term by setting $\kappa=0$. The first
term is:
\begin{align*}
\frac{1}{4}k^{2}\mathcal{A}\int_{0}^{\epsilon_{D}}\frac{d\varepsilon}{2\pi}\frac{1}{2}\frac{m^{2}}{\left(\varepsilon+\mu\right)^{3}}\frac{\tanh\left({\normalcolor \frac{\beta}{2}\varepsilon}\right)}{2\varepsilon} & =\frac{1}{32\pi}k^{2}\mathcal{A}m^{2}\int_{0}^{\epsilon_{D}}d\varepsilon\frac{\tanh\left(\frac{\beta}{2}\varepsilon\right)}{\varepsilon(\varepsilon+m)^{3}}\\
 & \approx\frac{1}{32\pi}k^{2}\mathcal{A}m^{2}\left[\int_{0}^{2/\beta}d\varepsilon\frac{\beta}{2(\varepsilon+m)^{3}}+\int_{2/\beta}^{\epsilon_{D}}d\varepsilon\frac{1}{\varepsilon(\varepsilon+m)^{3}}\right]\\
 & \approx\frac{1}{32\pi}k^{2}\mathcal{A}m^{2}\left\{ \frac{\beta}{4}\left[\frac{1}{m^{2}}-\frac{1}{(m+2T)^{2}}\right]\right.\\
 & \left.+\frac{1}{2m^{3}}\left[\frac{2m\epsilon_{D}}{\epsilon_{D}^{2}}+2\ln\left(\frac{\epsilon_{D}}{\epsilon_{D}+m}\right)-\frac{m(3m+4T)}{(m+2T)^{2}}-2\ln\left(\frac{2T}{m+2T}\right)\right]\right\} \\
 & \approx\frac{k^{2}\mathcal{A}}{64\pi m}\left[-1-2\ln2+2\ln\left(\beta m\right)\right].
\end{align*}
The latter term is:
\begin{align*}
\frac{1}{4}k^{2}\mathcal{A}\int_{0}^{\epsilon_{D}}\frac{d\varepsilon}{2\pi}\frac{1}{2}\frac{\tanh\left({\normalcolor \frac{\beta}{2}\varepsilon}\right)}{2\varepsilon\left(\varepsilon+\mu\right)} & =\frac{1}{32\pi}k^{2}\mathcal{A}\int_{0}^{\epsilon_{D}}d\varepsilon\ \frac{1}{\mu}\left(\frac{1}{\varepsilon}-\frac{1}{\varepsilon+\mu}\right)\tanh\left(\frac{\beta}{2}\varepsilon\right)\\
 & \approx\frac{1}{32\pi}k^{2}\mathcal{A}\int_{0}^{\epsilon_{D}}d\varepsilon\ \frac{1}{\mu}\frac{1}{\varepsilon}\tanh\left(\frac{\beta}{2}\varepsilon\right)-\frac{1}{16\pi}k^{2}\mathcal{A}\int_{0}^{\epsilon_{D}}d\varepsilon\ \frac{1}{\mu}\frac{1}{\varepsilon+\mu}\\
 & \approx\frac{1}{32\pi}k^{2}\mathcal{A}\frac{1}{m}\left[\ln\left(\frac{2e^{\gamma}}{\pi}\beta\epsilon_{D}\right)-\ln\left(1+\frac{\epsilon_{D}}{m}\right)\right]\\
 & \approx\frac{k^{2}\mathcal{A}}{32\pi m}\left[\ln\left(\frac{2e^{\gamma}}{\pi}\right)+\ln\left(\beta m\right)\right].
\end{align*}
Thus the total contribution at $\kappa=0$ (namely $\mu=m$) is:
\begin{equation}\label{supp_eq:chi_qm}
\chi_{\mathrm{qm}}(\kappa=0)k^{2}=\frac{k^{2}\mathcal{A}}{64\pi m}\left[-1+2\ln\left(\frac{e^{\gamma}}{\pi}\right)+4\ln\left(\beta m\right)\right].
\end{equation}

For conventional contribution, we have:
\begin{align}
\chi_{2,\mathrm{con}}k^{2} & =\mathcal{A}\int\frac{d^{2}\mathbf{q}}{(2\pi)^{2}}\left\{ \frac{\tanh\left[\frac{\beta}{2}\varepsilon\left(\mathbf{q}\right)\right]}{2\varepsilon\mathbf{(q})}-\frac{\tanh\left[\frac{\beta}{2}\varepsilon\left(\mathbf{q}+\frac{\mathbf{k}}{2}\right)\right]+\tanh\left[\frac{\beta}{2}\varepsilon\left(\mathbf{q}-\frac{\mathbf{k}}{2}\right)\right]}{2\left[\varepsilon\left(\mathbf{q}+\frac{\mathbf{k}}{2}\right)+\varepsilon\left(\mathbf{q}-\frac{\mathbf{k}}{2}\right)\right]}\right\}\nonumber \\
 & =\mathcal{A}\int\frac{d^{2}\mathbf{q}}{(2\pi)^{2}}\frac{1}{2}\left\{ \frac{1-2n_{F}\left[\varepsilon\left(\mathbf{q}\right)\right]}{2\varepsilon(\mathbf{q})}+\frac{n_{F}\left[\varepsilon\left(\mathbf{q}+\frac{\mathbf{k}}{2}\right)\right]+n_{F}\left[\varepsilon\left(\mathbf{q}-\frac{\mathbf{k}}{2}\right)\right]-1}{2\varepsilon(\mathbf{q})}\right\}\nonumber \\
 & \approx\mathcal{A}\int\frac{d^{2}\mathbf{q}}{(2\pi)^{2}}\frac{1}{2\varepsilon(\mathbf{q})}\partial_{\varepsilon}^{2}n_{F}[\varepsilon(\mathbf{q})]\left\{ \frac{v^{2}(\mathbf{q}\cdot\mathbf{k})}{2[\varepsilon(\mathbf{q})+\mu]}\right\} ^{2}\nonumber\\
 & \approx\mathcal{A}\int\frac{qdq}{2\pi}\frac{1}{2\varepsilon(\mathbf{q})}\partial_{\varepsilon}^{2}n_{F}[\varepsilon(\mathbf{q})]\frac{v^{4}q^{2}k^{2}}{8[\varepsilon(\mathbf{q})+\mu]^{2}}\nonumber\\
 & =\mathcal{A}\int_{-\kappa\epsilon_{D}}^{\epsilon_{D}}\frac{d\varepsilon}{32\pi}\frac{1}{\varepsilon}\partial_{\varepsilon}^{2}n_{F}(\varepsilon)\frac{k^{2}}{\varepsilon+\mu}\left[(\varepsilon+\mu)^{2}-m^{2}\right]\nonumber\\
 & =k^{2}\mathcal{A}\int_{-\kappa\epsilon_{D}}^{\epsilon_{D}}\frac{d\varepsilon}{32\pi}\partial_{\varepsilon}^{2}n_{F}(\varepsilon)\left[1+\frac{\mu}{\varepsilon}-\frac{m^{2}}{\varepsilon(\varepsilon+\mu)}\right]\nonumber\\
 & =k^{2}\mathcal{A}\int_{-\kappa\epsilon_{D}}^{\epsilon_{D}}\frac{d\varepsilon}{32\pi}\partial_{\varepsilon}^{2}n_{F}(\varepsilon)\left[1+\frac{\mu}{\varepsilon}\left(1-\frac{m^{2}}{\mu^{2}}\right)+\frac{m^{2}}{\mu(\varepsilon+\mu)}\right].
\end{align}

For case where $\kappa\beta\epsilon_{D}\gg1$, we have:
\begin{align}
\chi_{2,\mathrm{con}}k^{2} & \approx k^{2}\mathcal{A}\int_{-\kappa\epsilon_{D}}^{\epsilon_{D}}\frac{d\varepsilon}{32\pi}\partial_{\varepsilon}^{2}n_{F}(\varepsilon)\left[1+\frac{\mu}{\varepsilon}\left(1-\frac{m^{2}}{\mu^{2}}\right)+\frac{m^{2}}{\mu(\varepsilon+\mu)}\right]\nonumber \\
 & \approx k^{2}\mathcal{A}\int_{-\kappa\epsilon_{D}}^{\epsilon_{D}}\frac{d\varepsilon}{32\pi}\partial_{\varepsilon}^{2}n_{F}(\varepsilon)+k^{2}\mu\mathcal{A}\left(1-\frac{m^{2}}{\mu^{2}}\right)\int_{-\kappa\epsilon_{D}}^{\epsilon_{D}}\frac{d\varepsilon}{32\pi}\frac{\partial_{\varepsilon}^{2}n_{F}(\varepsilon)}{\varepsilon}\nonumber \\
 & \approx k^{2}\mathcal{A}\frac{1}{32\pi}\beta\left(e^{-\kappa\beta\epsilon_{D}}-e^{-\beta\epsilon_{D}}\right)+k^{2}\mu\mathcal{A}\frac{1}{32\pi}\left(1-\frac{m^{2}}{\mu^{2}}\right)\int_{-\infty}^{\infty}d\varepsilon\frac{\partial_{\varepsilon}^{2}n_{F}(\varepsilon)}{\varepsilon}\nonumber \\
 & \approx k^{2}\mu\mathcal{A}\frac{1}{32\pi}\left(1-\frac{m^{2}}{\mu^{2}}\right)\beta^{2}\frac{7\zeta(3)}{2\pi^{2}}\nonumber \\
 & \approx k^{2}\mathcal{A}\frac{7\zeta(3)}{64\pi^{3}}\frac{\kappa\epsilon_{D}}{T^{2}}.
\end{align}

For case where $|\kappa|\beta\epsilon_{D}\ll1$, we have:
\begin{align}
\chi_{2,\mathrm{con}}k^{2} & \approx k^{2}\mathcal{A}\int_{-\kappa\epsilon_{D}}^{\epsilon_{D}}\frac{d\varepsilon}{32\pi}\partial_{\varepsilon}^{2}n_{F}(\varepsilon)\left[1+\frac{\mu}{\varepsilon}\left(1-\frac{m^{2}}{\mu^{2}}\right)+\frac{m^{2}}{\mu(\varepsilon+\mu)}\right]\nonumber\\
 & \approx k^{2}\mathcal{A}\int_{-\kappa\epsilon_{D}}^{\epsilon_{D}}\frac{d\varepsilon}{32\pi}\partial_{\varepsilon}^{2}n_{F}(\varepsilon)\left(1+\frac{m}{\varepsilon}\frac{2\kappa\epsilon_{D}}{m}+\frac{m}{\varepsilon+m+\kappa\epsilon_{D}}\right)\nonumber\\
 & \approx\frac{k^{2}\mathcal{A}}{32\pi}\left[\int_{-\kappa\epsilon_{D}}^{\epsilon_{D}}d\varepsilon\ \partial_{\varepsilon}^{2}n_{F}(\varepsilon)+2\kappa\epsilon_{D}\int_{0}^{\epsilon_{D}}d\varepsilon\ \frac{\partial_{\varepsilon}^{2}n_{F}(\varepsilon)}{\varepsilon}+m\int_{-\kappa\epsilon_{D}}^{\epsilon_{D}}d\varepsilon\ \frac{\partial_{\varepsilon}^{2}n_{F}(\varepsilon)}{\varepsilon+m+\kappa\epsilon_{D}}\right]\nonumber\\
 & \approx\frac{k^{2}\mathcal{A}}{32\pi}\left[\beta\frac{e^{-\kappa\beta\epsilon_{D}}}{(1+e^{-\kappa\beta\epsilon_{D}})^{2}}+2\kappa\epsilon_{D}\beta^{2}\frac{7\zeta(3)}{4\pi^{2}}+\beta\frac{e^{-\kappa\beta\epsilon_{D}}}{(1+e^{-\kappa\beta\epsilon_{D}})^{2}}\right]\nonumber\\
 & \approx\frac{k^{2}\mathcal{A}}{32\pi}\left[\frac{\beta}{2}+2\kappa\epsilon_{D}\beta^{2}\frac{7\zeta(3)}{4\pi^{2}}\right]\nonumber\\
 & \approx\frac{k^{2}\mathcal{A}}{64\pi T}\left[1+\kappa\beta\epsilon_{D}\frac{7\zeta(3)}{\pi^{2}}\right].
\end{align}
As for the mean-field temperature, we have:
\begin{align}
\frac{1}{U} & =\mathcal{A}\int_{-\kappa\epsilon_{D}}^{\epsilon_{D}}\frac{(\varepsilon+\mu)d\varepsilon}{2\pi v^{2}}\frac{1}{2\varepsilon}\tanh\frac{\beta_{\mathrm{MF}}\varepsilon}{2}\nonumber\\
 & =\frac{\mathcal{A}}{4\pi v^{2}}\int_{-\kappa\epsilon_{D}}^{\epsilon_{D}}d\varepsilon\left(\tanh\frac{\beta_{\mathrm{MF}}\varepsilon}{2}+\frac{\mu}{\varepsilon}\tanh\frac{\beta_{\mathrm{MF}}\varepsilon}{2}\right)\nonumber\\
 & =\frac{\mathcal{A}}{4\pi v^{2}}\left[\frac{2}{\beta_{MF}}\ln\frac{\cosh(\beta_{\mathrm{MF}}\epsilon_{D}/2)}{\cosh(\kappa\beta_{\mathrm{MF}}\epsilon_{D}/2)}+\mu\int_{-\kappa\epsilon_{D}}^{\epsilon_{D}}d\varepsilon\ \frac{1}{\varepsilon}\tanh\frac{\beta_{\mathrm{MF}}\varepsilon}{2}\right].
\end{align}

Finally we should calculate $U-U^{2}\chi_{0}$:
\begin{align}
U-U^{2}\chi_{0} & =-U^{2}\mathcal{A}\int\frac{d^{2}\mathbf{q}}{(2\pi)^{2}}\frac{\tanh\left[\frac{\beta}{2}\varepsilon(\mathbf{q})\right]-\tanh\left[\frac{\beta_{\mathrm{MF}}}{2}\varepsilon(\mathbf{q})\right]}{2\varepsilon(\mathbf{q})}\nonumber \\
 & \approx-U^{2}\mathcal{A}\int\frac{d^{2}\mathbf{q}}{(2\pi)^{2}}\frac{\mathrm{sech}^{2}\left[\frac{\beta_{\mathrm{MF}}}{2}\varepsilon(\mathbf{q})\right]}{2\varepsilon(\mathbf{q})}\frac{\beta-\beta_{\mathrm{MF}}}{2}\varepsilon(\mathbf{q})\nonumber \\
 & =-U^{2}\mathcal{A}\frac{1}{v^{2}}\int_{-\kappa\epsilon_{D}}^{\epsilon_{D}}\frac{d\varepsilon}{2\pi}(\varepsilon+\mu)\frac{\mathrm{sech}^{2}\left(\frac{\beta_{\mathrm{MF}}\varepsilon}{2}\right)}{4}(\beta-\beta_{\mathrm{MF}})\nonumber \\
 & =-U^{2}\mathcal{A}\frac{1}{8\pi v^{2}}(\beta-\beta_{\mathrm{MF}})\frac{2}{\beta_{\mathrm{MF}}}\left\{ \left.\left[(\varepsilon+m+\kappa\epsilon_{D})\tanh\frac{\beta_{\mathrm{MF}}\varepsilon}{2}\right]\right|_{-\kappa\epsilon_{D}}^{\epsilon_{D}}-\int_{-\kappa\epsilon_{D}}^{\epsilon_{D}}d\varepsilon\tanh\frac{\beta_{\mathrm{MF}}\varepsilon}{2}\right\} \nonumber \\
 & =\frac{U^{2}\mathcal{A}}{4\pi v^{2}}\frac{\beta_{\mathrm{MF}}-\beta}{\beta_{\mathrm{MF}}}\left\{ \left[m+(1+\kappa)\epsilon_{D}\right]\tanh\frac{\beta_{\mathrm{MF}}\epsilon_{D}}{2}+m\tanh\frac{\kappa\beta_{\mathrm{MF}}\epsilon_{D}}{2}\right.\label{eq:U-U2X}\nonumber\\
 & \left.-\frac{2}{\beta_{MF}}\left(\ln\cosh\frac{\beta_{\mathrm{MF}}\epsilon_{D}}{2}-\ln\cosh\frac{\kappa\beta_{\mathrm{MF}}\epsilon_{D}}{2}\right)\right\},
\end{align}
where for convenience we have defined $\alpha=4\pi v^{2}/U\mathcal{A}$.
For $\kappa\beta\epsilon_{D}\gg1$, we have:
\begin{align}
\alpha & \approx\frac{2}{\beta_{MF}}\ln\frac{\cosh(\beta_{\mathrm{MF}}\epsilon_{D}/2)}{\cosh(\kappa\beta_{\mathrm{MF}}\epsilon_{D}/2)}+\mu\int_{-\kappa\epsilon_{D}}^{\epsilon_{D}}d\varepsilon\ \frac{1}{\varepsilon}\tanh\frac{\beta_{\mathrm{MF}}\varepsilon}{2}\nonumber\\
 & \approx(1-\kappa)\epsilon_{D}+2(m+\kappa\epsilon_{D})\ln\left(\frac{2e^{\gamma}}{\pi}\beta_{\mathrm{MF}}\epsilon_{D}\right)+(m+\kappa\epsilon_{D})\ln\kappa\nonumber\\
T_{\mathrm{MF}} & \approx\frac{2\sqrt{\kappa}}{\pi}\epsilon_{D}\exp\left(\gamma+\frac{1-\kappa}{2\kappa}-\frac{\alpha}{2\kappa\epsilon_{D}}\right).
\end{align}
For $\kappa\beta\epsilon_{D}\ll1$, we have:
\begin{align}
\alpha & \approx\frac{2}{\beta_{MF}}\ln\frac{\cosh(\beta_{\mathrm{MF}}\epsilon_{D}/2)}{\cosh(\kappa\beta_{\mathrm{MF}}\epsilon_{D}/2)}+\mu\int_{-\kappa\epsilon_{D}}^{\epsilon_{D}}d\varepsilon\ \frac{1}{\varepsilon}\tanh\frac{\beta_{\mathrm{MF}}\varepsilon}{2}\nonumber\\
 & \approx\epsilon_{D}+(m+\kappa\epsilon_{D})\ln\left(\frac{2e^{\gamma}}{\pi}\beta_{\mathrm{MF}}\epsilon_{D}\right)+(m+\kappa\epsilon_{D})\frac{\kappa\beta_{\mathrm{MF}}\epsilon_{D}}{2}\nonumber\\
 & \approx\epsilon_{D}+(m+\kappa\epsilon_{D})\ln\left(\frac{2e^{\gamma}}{\pi}\beta_{\mathrm{MF}}\epsilon_{D}\right)\nonumber\\
T_{\mathrm{MF}} & \approx\frac{2}{\pi}\epsilon_{D}\exp\left(\gamma+\frac{\epsilon_{D}-\alpha}{\mu}\right).
\end{align}
We should also calculate $U-U^{2}\chi_{0}$ to determine the coherence
length:
\begin{align}
U-U^{2}\chi_{0} & =-U^{2}\mathcal{A}\int\frac{d^{2}\mathbf{q}}{(2\pi)^{2}}\frac{\tanh\left[\frac{\beta}{2}\varepsilon(\mathbf{q})\right]-\tanh\left[\frac{\beta_{\mathrm{MF}}}{2}\varepsilon(\mathbf{q})\right]}{2\varepsilon(\mathbf{q})}\nonumber \\
 & \approx-U^{2}\mathcal{A}\int\frac{d^{2}\mathbf{q}}{(2\pi)^{2}}\frac{\mathrm{sech}^{2}\left[\frac{\beta_{\mathrm{MF}}}{2}\varepsilon(\mathbf{q})\right]}{2\varepsilon(\mathbf{q})}\frac{\beta-\beta_{\mathrm{MF}}}{2}\varepsilon(\mathbf{q})\nonumber \\
 & =-U^{2}\mathcal{A}\frac{1}{v^{2}}\int_{-\kappa\epsilon_{D}}^{\epsilon_{D}}\frac{d\varepsilon}{2\pi}(\varepsilon+\mu)\frac{\mathrm{sech}^{2}\left(\frac{\beta_{\mathrm{MF}}\varepsilon}{2}\right)}{4}(\beta-\beta_{\mathrm{MF}})\nonumber \\
 & =-U^{2}\mathcal{A}\frac{1}{8\pi v^{2}}(\beta-\beta_{\mathrm{MF}})\frac{2}{\beta_{\mathrm{MF}}}\left\{ \left.\left[(\varepsilon+m+\kappa\epsilon_{D})\tanh\frac{\beta_{\mathrm{MF}}\varepsilon}{2}\right]\right|_{-\kappa\epsilon_{D}}^{\epsilon_{D}}-\int_{-\kappa\epsilon_{D}}^{\epsilon_{D}}d\varepsilon\tanh\frac{\beta_{\mathrm{MF}}\varepsilon}{2}\right\} \nonumber \\
 & =\frac{U^{2}\mathcal{A}}{4\pi v^{2}}\frac{\beta_{\mathrm{MF}}-\beta}{\beta_{\mathrm{MF}}}\left\{ \left[m+(1+\kappa)\epsilon_{D}\right]\tanh\frac{\beta_{\mathrm{MF}}\epsilon_{D}}{2}+m\tanh\frac{\kappa\beta_{\mathrm{MF}}\epsilon_{D}}{2}\right.\nonumber\\
 & \left.-\frac{2}{\beta_{MF}}\left(\ln\cosh\frac{\beta_{\mathrm{MF}}\epsilon_{D}}{2}-\ln\cosh\frac{\kappa\beta_{\mathrm{MF}}\epsilon_{D}}{2}\right)\right\}.
\end{align}

Specifying to the case where $\epsilon_{D}\gg m,\ \kappa\beta_{\mathrm{MF}}\epsilon_{D}\gg1$,
we have:
\begin{align}
U-U^{2}\chi_{0} & \approx\frac{U}{\alpha}\frac{\beta_{\mathrm{MF}}-\beta}{\beta_{\mathrm{MF}}}\left\{ (1+\kappa)\epsilon_{D}+m-\frac{2}{\beta_{MF}}\left(\frac{\beta_{\mathrm{MF}}\epsilon_{D}}{2}-\frac{\kappa\beta_{\mathrm{MF}}\epsilon_{D}}{2}\right)\right\}\nonumber \\
 & \approx2\kappa\epsilon_{D}\frac{U}{\alpha}\frac{\beta_{\mathrm{MF}}-\beta}{\beta_{\mathrm{MF}}}.
\end{align}
For $\kappa\beta\epsilon_{D}\ll1$, we have:
\begin{align}
U-U^{2}\chi_{0} & \approx\frac{U}{\alpha}\frac{\beta_{\mathrm{MF}}-\beta}{\beta_{\mathrm{MF}}}\left\{ \left[m+(1+\kappa)\epsilon_{D}\right]-\frac{2}{\beta_{MF}}\left(\frac{\beta_{\mathrm{MF}}\epsilon_{D}}{2}-\ln2\right)\right\} \nonumber\\
 & \approx \mu\frac{U}{\alpha}\frac{\beta_{\mathrm{MF}}-\beta}{\beta_{\mathrm{MF}}}.
\end{align}
Note that we can write down an effective Lagrangian for our bosonic
field:
\begin{align}
\mathcal{L}[\delta\Delta,\delta\bar{\Delta}] & =\int\frac{d^{2}\mathbf{k}}{(2\pi)^{2}}\ F_{2}(\mathbf{k})\nonumber \\
 & =(U-U^{2}\chi_{0})|\delta\Delta|^{2}+(\chi_{2,\mathrm{qm}}+\chi_{2,\mathrm{con}})|\partial_{\mathbf{k}}\delta\Delta|^{2}.
\end{align}
As such, the coherence length should be given by:
\begin{align}
\xi & =\sqrt{-U^{2}\frac{\chi_{2,\mathrm{qm}}+\chi_{2,\mathrm{con}}}{U-U^{2}\chi_{0}}}\nonumber \\
 & =\sqrt{-\frac{U^{2}\chi_{2,\mathrm{qm}}}{U-U^{2}\chi_{0}}-\frac{U\chi_{2,\mathrm{con}}}{U-U^{2}\chi_{0}}}\nonumber \\
 & =\sqrt{\xi_{\mathrm{qm}}^{2}+\xi_{\mathrm{con}}^{2}}\label{eq:lc}.
\end{align}
At the $\kappa\beta_{\mathrm{MF}}\epsilon_{D}\gg1$ regime, the coherence
lengths are given by:
\begin{align}
\xi_{\mathrm{qm}} & =\sqrt{-\frac{U^{2}\chi_{2,\mathrm{qm}}}{U-U^{2}\chi_{0}}}\nonumber\\
 & \approx\sqrt{-\frac{\frac{U^{2}\mathcal{A}}{32\pi\kappa\epsilon_{D}}\left[\ln\left(\frac{\epsilon_{D}^{3}}{mT_{\mathrm{MF}}^{2}}\right)+2\ln\left(\frac{2e^{\gamma}}{\pi}\right)+\ln\left(\frac{\kappa^{3}}{1+\kappa}\right)\right]}{2\kappa\epsilon_{D}\frac{U^{2}\mathcal{A}}{4\pi v^{2}}\frac{\beta_{\mathrm{MF}}-\beta}{\beta_{\mathrm{MF}}}}}\nonumber\\
 & =\frac{v}{4\mu}\left(\frac{T_{\mathrm{MF}}-T}{T_{\mathrm{MF}}}\right)^{-1/2}\sqrt{\ln\left(\frac{\mu^{3}}{mT_{\mathrm{MF}}^{2}}\right)}\\
\xi_{\mathrm{con}} & =\sqrt{-\frac{U^{2}\chi_{2,\mathrm{con}}}{U-U^{2}\chi_{0}}}\nonumber\\
 & \approx\sqrt{-\frac{U^{2}\mathcal{A}\frac{7\zeta(3)}{64\pi^{3}}\frac{\kappa\epsilon_{D}}{T_{\mathrm{MF}}^{2}}}{2\kappa\epsilon_{D}\frac{U^{2}\mathcal{A}}{4\pi v^{2}}\frac{\beta_{\mathrm{MF}}-\beta}{\beta_{\mathrm{MF}}}}}\nonumber\\
 & =\frac{v}{T_{\mathrm{MF}}}\left(\frac{T_{\mathrm{MF}}-T}{T_{\mathrm{MF}}}\right)^{-1/2}\sqrt{\frac{7\zeta(3)}{32\pi^{2}}}.
\end{align}

As such, the conventional contribution dominates, by a factor of $\epsilon_{D}/T_{\mathrm{MF}}\gg1$.
For $\kappa=0$, the coherence lengths are given by:
\begin{align}
\xi_{\mathrm{qm}} & =\sqrt{-\frac{U^{2}\chi_{2,\mathrm{qm}}}{U-U^{2}\chi_{0}}}\nonumber\\
 & \approx\sqrt{-\frac{\frac{U^{2}\mathcal{A}}{64\pi m}\left[-1+2\ln\left(\frac{e^{\gamma}}{\pi}\right)+4\ln\left(\beta_{\mathrm{MF}}m\right)\right]}{m\frac{U^{2}\mathcal{A}}{4\pi v^{2}}\frac{\beta_{\mathrm{MF}}-\beta}{\beta_{\mathrm{MF}}}}}\nonumber\\
 & =\frac{v}{4m}\left(\frac{T_{\mathrm{MF}}-T}{T_{\mathrm{MF}}}\right)^{-1/2}\sqrt{-1+2\ln\left(\frac{e^{\gamma}}{\pi}\right)+4\ln\left(\beta_{\mathrm{MF}}m\right)}\\
\xi_{\mathrm{con}} & =\sqrt{-\frac{U^{2}\chi_{2,\mathrm{con}}}{U-U^{2}\chi_{0}}}\nonumber\\
 & \approx\sqrt{-\frac{\frac{U^{2}\mathcal{A}}{64\pi T_{\mathrm{MF}}}}{m\frac{U^{2}\mathcal{A}}{4\pi v^{2}}\frac{\beta_{\mathrm{MF}}-\beta}{\beta_{\mathrm{MF}}}}}\nonumber\\
 & =\frac{1}{4}\frac{v}{\sqrt{mT_{\mathrm{MF}}}}\left(\frac{T_{\mathrm{MF}}-T}{T_{\mathrm{MF}}}\right)^{-1/2}.
\end{align}
Note that in this case the quantum metric contribution still smaller, but is comparable with the conventional contribution by a factor of $2\sqrt{\ln(\beta_\mathrm{MF}m)/\beta_\mathrm{MF}m}$.

\subsubsection{Beyond band edge}
In this part, we aim to provide a brief discussion on the case beyond the band edge.
We assume $||\mu|-m|\ll m$ such that $\chi_{2,\mathrm{qm}}=\chi_{2,\mathrm{qm}}(\kappa=0)$, corresponding to $\mu=m$ with $\chi_{2,qm}(\kappa)$ defined in Eq.~\eqref{supp_eq:chi_qm}.
We should calculate $U-U^{2}\chi_{0}$ in the vicinity where superconductivity vanishes, namely 
when $|\beta_\mathrm{MF}(||\mu|-m|)|\gg 0$:

\begin{align}
    & \frac{U^{2}\mathcal{A}}{4\pi v^{2}}\frac{\beta_{\mathrm{MF}}-\beta}{\beta_{\mathrm{MF}}}\left[ \left(\mu+\epsilon_{D}\right)\tanh\frac{\beta_{\mathrm{MF}}\epsilon_{D}}{2}+m\tanh\frac{\beta_{\mathrm{MF}}(\mu-m)}{2}-\frac{2}{\beta_{MF}}\left(\ln\cosh\frac{\beta_{\mathrm{MF}}\epsilon_{D}}{2}+\ln\cosh\frac{\beta_{\mathrm{MF}}(\mu-m)}{2}\right)\right] \nonumber\\  
    \approx\ &\frac{U^{2}\mathcal{A}}{4\pi v^{2}}\frac{\beta_{\mathrm{MF}}-\beta}{\beta_{\mathrm{MF}}}\left[ \mu+\epsilon_{D}-m-\frac{2}{\beta_{MF}}\left(\frac{\beta_{\mathrm{MF}}\epsilon_{D}}{2}+\frac{\beta_{\mathrm{MF}}(\mu-m)}{2}\right)\right]+\frac{4}{\beta_\mathrm{MF}}\ln 2\nonumber\\
    =\ &\frac{U^{2}\mathcal{A}}{4\pi v^{2}}\frac{\beta_{\mathrm{MF}}-\beta}{\beta_{\mathrm{MF}}} 4T_\mathrm{MF}\ln 2.
\end{align}
which vanishes as the mean-field transition temperature goes to zero due to vanishing density of state. As such the quantum metric coherence length diverges at boundary of phase transition. Note that the conventional contribution remains finite until $T_\mathrm{MF}\rightarrow0$ due to both $\chi_{2,\mathrm{con}}\propto \beta_\mathrm{MF}e^{\kappa\beta_\mathrm{MF}\epsilon_D}$ and $U-U^2\chi_0$ vanishing as $T_\mathrm{MF}\rightarrow0$. It is important to note that the superconductivity exists due to a finite temperature effect, and is expected to vanish as temperature goes to 0.

\subsection{Case 2: $T_{\mathrm{MF}}\gg m$}
In this scenario, most of the calculation is identical to the previous
case, with the exception of $\chi_{\mathrm{qm}}$ when $\mu\approx m$:
\begin{align}
\frac{1}{4}k^{2}\mathcal{A}\int_{0}^{\epsilon_{D}}\frac{d\varepsilon}{2\pi}\frac{1}{2}\frac{\tanh\left({\normalcolor \frac{\beta}{2}\varepsilon}\right)}{2\varepsilon\left(\varepsilon+\mu\right)} & =\frac{1}{32\pi}k^{2}\mathcal{A}\int_{0}^{\epsilon_{D}}d\varepsilon\ \frac{1}{\mu}\left(\frac{1}{\varepsilon}-\frac{1}{\varepsilon+\mu}\right)\tanh\left(\frac{\beta}{2}\varepsilon\right)\nonumber\\
 & =\frac{1}{32\pi}k^{2}\mathcal{A}\int_{0}^{\epsilon_{D}}d\varepsilon\ \frac{1}{\mu}\frac{1}{\varepsilon}\tanh\left(\frac{\beta}{2}\varepsilon\right)-\frac{1}{16\pi}k^{2}\mathcal{A}\int_{0}^{\epsilon_{D}}d\varepsilon\ \frac{1}{\mu}\frac{1}{\varepsilon+\mu}\tanh\left(\frac{\beta}{2}\varepsilon\right)\nonumber\\
 & \approx\frac{1}{32\pi}k^{2}\mathcal{A}\frac{1}{m}\left[\int_{0}^{\epsilon_{D}}d\varepsilon\ \frac{1}{\varepsilon}\tanh\left(\frac{\beta}{2}\varepsilon\right)-\int_{m}^{\epsilon_{D}}d\varepsilon\ \frac{1}{\varepsilon}\left(1-\frac{m}{\varepsilon}\right)\tanh\left(\frac{\beta}{2}\varepsilon\right)\right]\\
 & =\frac{1}{32\pi}k^{2}\mathcal{A}\int_{m}^{\epsilon_{D}}d\varepsilon\ \frac{1}{\varepsilon^{2}}\tanh\left(\frac{\beta}{2}\varepsilon\right)\nonumber\\
 & \approx\frac{k^{2}\mathcal{A}}{32\pi}\left[\int_{m}^{2/\beta}d\varepsilon\ \frac{\beta}{2\varepsilon}+\int_{2/\beta}^{\epsilon_{D}}d\varepsilon\ \frac{1}{\varepsilon^{2}}\right]\nonumber\\
 & \approx\frac{k^{2}\mathcal{A}}{32\pi}\left[\frac{\beta}{2}\ln\frac{2}{\beta m}-\frac{1}{\epsilon_{D}}+\frac{1}{2T}\right]\nonumber\\
 & \approx\frac{k^{2}\mathcal{A}}{64\pi T}\left(1-\ln\frac{\beta m}{2}\right).
\end{align}

\begin{figure}[t]
    \centering
    \includegraphics[width=1\columnwidth]{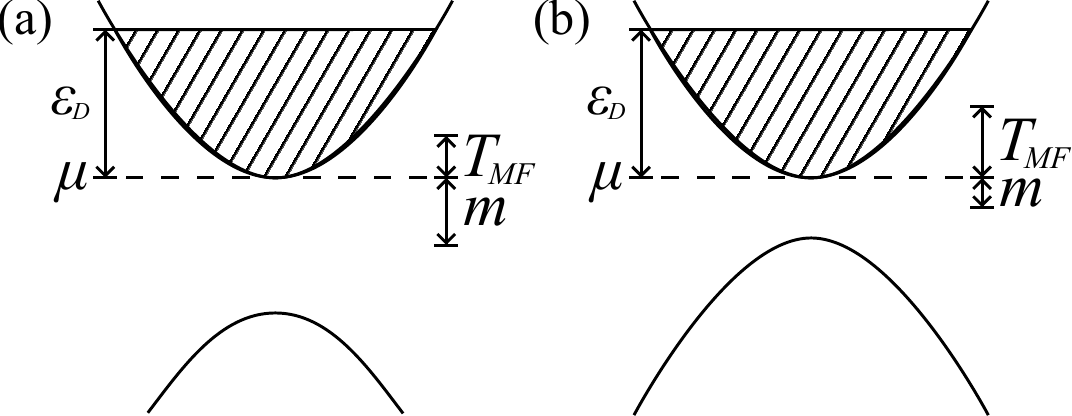}
    \caption{Illustration of the energy scale and band structure for the toy model. 
    The dashed area denote the energy range ($\epsilon_D$) where the interaction is effectively attractive.
    In the figure we have assumed the chemical potential $\mu=m$, and $\epsilon_D\gg m,\ T_\mathrm{MF}$. Figure (a) correspond to the case where $m\gg T_\mathrm{MF}$, while figure (b) correspond to the case where $m\ll T_\mathrm{MF}$} \label{fig: BS}
    \label{fig:Toy_model}
\end{figure}

As such, the coherence length at $\kappa=0$ should change accordingly:
\begin{align}
\xi_{\mathrm{qm}} & =\sqrt{-\frac{U^{2}\chi_{2,\mathrm{qm}}}{U-U^{2}\chi_{0}}}\nonumber\\
 & \approx\sqrt{-\frac{\frac{U^{2}\mathcal{A}}{64\pi T_{\mathrm{MF}}}\left(1-\ln\frac{\beta_{\mathrm{MF}}m}{2}\right)}{m\frac{U^{2}\mathcal{A}}{4\pi v^{2}}\frac{\beta_{\mathrm{MF}}-\beta}{\beta_{\mathrm{MF}}}}}\nonumber\\
 & =\frac{1}{4}\frac{v}{\sqrt{mT_{\mathrm{MF}}}}\left(\frac{T_{\mathrm{MF}}-T}{T_{\mathrm{MF}}}\right)^{-1/2}\sqrt{1-\ln\frac{\beta_{\mathrm{MF}}m}{2}},\\
\xi_{\mathrm{con}} & =\sqrt{-\frac{U^{2}\chi_{2,\mathrm{con}}}{U-U^{2}\chi_{0}}}\nonumber\\
 & \approx\sqrt{-\frac{\frac{U^{2}\mathcal{A}}{64\pi T_{\mathrm{MF}}}}{m\frac{U^{2}\mathcal{A}}{4\pi v^{2}}\frac{\beta_{\mathrm{MF}}-\beta}{\beta_{\mathrm{MF}}}}}\nonumber\\
 & =\frac{1}{4}\frac{v}{\sqrt{mT_{\mathrm{MF}}}}\left(\frac{T_{\mathrm{MF}}-T}{T_{\mathrm{MF}}}\right)^{-1/2}.
\end{align}
Note the two contributions share a similar form, with the quantum
metric contribution having an extra factor of $\sqrt{1-\ln(\beta_{\mathrm{MF}}m/2)}>1$. The quantum metric, thus plays a dominating role in this regime, when
superconductivity occurs near the band edge. The results for the toy model is summarized in Table~\ref{Table_TM}, alongside an illustration of the energy scale and band structure in Fig.~\ref{fig: BS}.
\section{Discussion on pairing and time-reversal symmetry in IVC}

In this section, we aim to discuss the pairing of the IVC state, in
order to induce superconductivity. To begin with, we can write down
our IVC state in the valley basis $\begin{pmatrix}K\\
K'
\end{pmatrix}$:

\begin{align}
\psi_{-,\mathbf{q}} & =\frac{1}{\sqrt{2}}
\begin{pmatrix}
e^{-i\phi_{\mathbf{q}}}\sqrt{1-\frac{\Delta_{z}(\mathbf{q})}{|\boldsymbol{\Delta}(\mathbf{q})|}}\\
\sqrt{1+\frac{\Delta_{z}(\mathbf{q})}{|\boldsymbol{\Delta}(\mathbf{q})|}}
\end{pmatrix},\\
\psi_{-,\mathbf{-q}} & =\frac{1}{\sqrt{2}}\begin{pmatrix}e^{-i\phi_{-\mathbf{q}}}\sqrt{1-\frac{\Delta_{z}(-\mathbf{q})}{|\boldsymbol{\Delta}(-\mathbf{q})|}}\\
\sqrt{1+\frac{\Delta_{z}(-\mathbf{q})}{|\boldsymbol{\Delta}(-\mathbf{q})|}}
\end{pmatrix}\nonumber\\
 & =\frac{1}{\sqrt{2}}\begin{pmatrix}e^{-i\phi_{\mathbf{q}}}\sqrt{1+\frac{\Delta_{z}(\mathbf{q})}{|\boldsymbol{\Delta}(\mathbf{q})|}}\\
\sqrt{1-\frac{\Delta_{z}(\mathbf{q})}{|\boldsymbol{\Delta}(\mathbf{q})|}}
\end{pmatrix}.
\end{align}

Recall the IVC order parameter's phase can be matched with the phase
of the inter-valley form factor $\left\langle u_{K,\mathbf{k}}\right.\left|u_{K',\mathbf{k}}\right\rangle $\cite{IVC},
inducing a six-fold symmetry in the phase, resulting in $\phi_{\mathbf{q}}=\phi_{-\mathbf{q}}$.
As such the form factor is given by:
\begin{align}
\Gamma(\mathbf{q},0) & =\psi_{-,\mathbf{-q}}(K)\psi_{-,\mathbf{q}}(K')+\psi_{-,\mathbf{-q}}(K')\psi_{-,\mathbf{q}}(K)\nonumber\\
 & =\psi_{-,\mathbf{-q}}^{T}\sigma_{x}\psi_{-,\mathbf{q}}\nonumber\\
 & =\frac{1}{2}\left[\left(1+\frac{\Delta_{z}}{|\boldsymbol{\Delta}|}\right)e^{-i\phi_{\mathbf{q}}}+\left(1-\frac{\Delta_{z}}{|\boldsymbol{\Delta}|}\right)e^{-i\phi_{\mathbf{q}}}\right]\nonumber\\
 & =e^{-i\phi_{\mathbf{q}}}\\
|\Gamma(\mathbf{q},0)|^{2} & =1.
\end{align}
\renewcommand{\arraystretch}{1.8}
\begin{table}[h]
\begin{centering}\label{tab-2}
\caption{Summary on the toy model, where $\frac{1}{U'}=\frac{1}{U}-\frac{\epsilon_{D}\mathcal{A}}{4\pi v^{2}}$}
\begin{tabular}{|c|c|c|c|c|}
\hline 
 &$|\mu|$&$T_\mathrm{MF}$&$\xi_\mathrm{con}$& $\xi_\mathrm{qm}$\tabularnewline
\hline 
$m\gg T_\mathrm{MF}$
&\multirow{2}{*}{$\gg m$}
&\multirow{2}{*}{$\propto\sqrt{\rho\epsilon_{D}}\exp\left(-\frac{1}{\rho U'}\right)$}
&\multirow{2}{*}{$\propto\frac{v}{T_\mathbf{MF}}$}
&\multirow{3}{*}{$\propto\frac{v}{\mu}\sqrt{\ln{(\beta_\mathrm{MF}\mu})}$}
\tabularnewline
\cline{1-1}
$m\ll T_\mathrm{MF}$
&
&
&
&
\tabularnewline
\cline{1-4}
$m\gg T_\mathrm{MF}$
&\multirow{2}{*}{$\rightarrow m$}
&\multirow{2}{*}{$\propto\epsilon_{D}\exp\left(-\frac{2}{\rho U'}\right)$}
&\multirow{2}{*}{$\propto\frac{v}{\sqrt{m T_\mathrm{MF}}}$}
&
\tabularnewline
\cline{1-1}
\cline{5-5}
$m\ll T_\mathrm{MF}$
&
&
&
&$\propto\frac{v}{\sqrt{mT_\mathrm{MF}}}\sqrt{1-\ln\frac{\beta_{\mathrm{MF}}m}{2}}$
\tabularnewline
\hline
\end{tabular}\label{Table_TM}
\par\end{centering}
\end{table}
As such we have recovered time reversal invariant for the IVC state. With the invariant being intact after introducing the IVC order parameter, our previous calculation on the
conventional coherence length remains valid. As for the projector
matrices, we note the symmetry:
\begin{align}
P(\mathbf{q}) & =\psi_{-,\mathbf{q}}\psi_{-,\mathbf{q}}^{\dagger}\nonumber\\
 & =\frac{1}{2}\begin{pmatrix}e^{-i\phi_{\mathbf{q}}}\sqrt{1-\frac{\Delta_{z}}{|\boldsymbol{\Delta}|}}\\
\sqrt{1+\frac{\Delta_{z}}{|\boldsymbol{\Delta}|}}
\end{pmatrix}
\begin{pmatrix}e^{i\phi_{\mathbf{q}}}\sqrt{1-\frac{\Delta_{z}}{|\boldsymbol{\Delta}|}} & \sqrt{1+\frac{\Delta_{z}}{|\boldsymbol{\Delta}|}}\end{pmatrix}\nonumber\\
 & =\frac{1}{2}\begin{pmatrix}1-\frac{\Delta_{z}}{|\boldsymbol{\Delta}|} & e^{-i\phi_{\mathbf{q}}}\sqrt{1-\frac{\Delta_{z}^{2}}{|\boldsymbol{\Delta}|^{2}}}\\
e^{i\phi_{\mathbf{q}}}\sqrt{1-\frac{\Delta_{z}^{2}}{|\boldsymbol{\Delta}|^{2}}} & 1+\frac{\Delta_{z}}{|\boldsymbol{\Delta}|}
\end{pmatrix},\\
\tilde{P}(-\mathbf{q}) & =\psi_{-,-\mathbf{q}}^{*}\psi_{-,-\mathbf{q}}^{T}\nonumber\\
 & =\frac{1}{2}
 \begin{pmatrix}
 e^{i\phi_{-\mathbf{q}}}\sqrt{1+\frac{\Delta_{z}}{|\boldsymbol{\Delta}|}}\\
\sqrt{1-\frac{\Delta_{z}}{|\boldsymbol{\Delta}|}}
\end{pmatrix}
\begin{pmatrix}
e^{-i\phi_{-\mathbf{q}}}\sqrt{1+\frac{\Delta_{z}}{|\boldsymbol{\Delta}|}} 
& \sqrt{1-\frac{\Delta_{z}}{|\boldsymbol{\Delta}|}}
\end{pmatrix}\nonumber\\
 & =\frac{1}{2}\begin{pmatrix}1+\frac{\Delta_{z}}{|\boldsymbol{\Delta}|} & e^{i\phi_{\mathbf{q}}}\sqrt{1-\frac{\Delta_{z}^{2}}{|\boldsymbol{\Delta}|^{2}}}\\
e^{-i\phi_{\mathbf{q}}}\sqrt{1-\frac{\Delta_{z}^{2}}{|\boldsymbol{\Delta}|^{2}}} & 
1-\frac{\Delta_{z}}{|\boldsymbol{\Delta}|}
\end{pmatrix},\\
\sigma_{x}\tilde{P}(-\mathbf{q})\sigma_{x} & =\frac{1}{2}
\begin{pmatrix}
e^{-i\phi_{\mathbf{q}}}\sqrt{1-\frac{\Delta_{z}^{2}}{|\boldsymbol{\Delta}|^{2}}} & 1-\frac{\Delta_{z}}{|\boldsymbol{\Delta}|}\\
1+\frac{\Delta_{z}}{|\boldsymbol{\Delta}|} & e^{i\phi_{\mathbf{q}}}\sqrt{1-\frac{\Delta_{z}^{2}}{|\boldsymbol{\Delta}|^{2}}}
\end{pmatrix}\sigma_{x}\nonumber\\
 & =\frac{1}{2}\begin{pmatrix}1-\frac{\Delta_{z}}{|\boldsymbol{\Delta}|} & e^{-i\phi_{\mathbf{q}}}\sqrt{1-\frac{\Delta_{z}^{2}}{|\boldsymbol{\Delta}|^{2}}}\\
e^{i\phi_{\mathbf{q}}}\sqrt{1-\frac{\Delta_{z}^{2}}{|\boldsymbol{\Delta}|^{2}}} & 1+\frac{\Delta_{z}}{|\boldsymbol{\Delta}|}
\end{pmatrix}\nonumber\\
 & =P(\mathbf{q}).
\end{align}

Note that the expansion of form factor around $\mathbf{k}=0$ retrieves
the quantum metric:
\begin{align}
\left|\psi_{-,\mathbf{-q}}^{T}\sigma_{x}\psi_{-,\mathbf{q}+\mathbf{k}}\right|{}^{2} & =\mathrm{Tr}{\left[\sigma_{x}\tilde{P}(-\mathbf{q})\sigma_{x}P(\mathbf{q}+\mathbf{k})\right]}\nonumber\\
 & =\mathrm{Tr}{\left[P(\mathbf{q})P(\mathbf{q}+\mathbf{k})\right]}\nonumber\\
 & =1-\frac{1}{2}\sum_{i,j}k_{i}k_{j}\mathrm{Tr}[\partial_{i}P(\mathbf{q})\partial_{j}P(\mathbf{q})]\nonumber\\
 & =1-\sum_{i,j}k_{i}k_{j}g_{ij}(\mathbf{q}),
\end{align}
where the intermediate step is similar to \eqref{qm} thus are not shown explicitly. As such, our results obtained for quantum metric coherence length from bare electrons,
is also applicable for superconductivity originating from the IVC
state, besides a change in the Bloch state, thus the quantum metric.
This justifies the application of the Ginzburg-Landau theory developed in the toy model section, to study the superconductivity originating from the IVC state in RTG.

\section{Details of fitting procedure for $U$ and $\Delta_{\mathrm{IVC}}$}
In this section, we aim to provide details of the fitting procedures for the effective attractive interaction strength $U$ and the IVC order parameter $\Delta_{\mathrm{IVC}}$ for RTG.\\ \\
In section ``Discrepancies in superconductivity of RTG", the effective attractive interaction strength $U$ is obtained from the linearized gap equation \eqref{SC} assuming effective attractive interaction between quasiparticles \eqref{Hint}. To be precise, we projected onto the valance band at the Fermi energy of the flavor-symmetric 6-band Hamiltonian of RTG \cite{HM-RTG}, which is justified due to the band gap separating the conduction and valence bands. The effective attractive interaction strength $U=4$meV is fitted such that \eqref{SC} reproduced the maximal value of transition temperature observed experimentally  $T_c\sim 120$mK at the charge carrier density $n=-1.85\times 10^{12}\mathrm{cm}^{-2}$.
\\ \\
In section ``Numerical calculations on the microscopic model", the IVC order parameter $\Delta_\mathrm{IVC}$ is fitted using the IVC quasiparticle band structure. From the quasiparticle band structure, we identify the charge carrier density at the band edge, which corresponds to the phase boundary between the IVC-SC and IVC state. We then shift the phase boundary by tuning the IVC order parameter, until we can match the charge carrier density of $n\approx-1.8\times 10^{12}\mathrm{cm}^{-2}$, where a sharp disappearance of superconductivity is observed in the experiment. The fitting procedure for effective attractive interaction strength $U$ is similar to section ``Discrepancies in superconductivity of RTG". However since the effective attractive interaction is between IVC quasiparticles, we have to instead project onto the lower energy IVC quasiparticle band. Following the procedure previously discussed we obtain effective attractive interaction strength of $U=1.2$meV instead. 

\end{document}